\documentclass[apj,iop]{emulateapj}
\usepackage[utf8]{inputenc}
\usepackage{footnote}
\graphicspath{ {./Figures/} }
\usepackage{tabularx}
\usepackage[symbol]{footmisc}
\usepackage{float}
\usepackage{makecell}
\usepackage{placeins}
\usepackage{threeparttable}
\usepackage{natbib}
\usepackage{amsmath}
\usepackage[caption=false]{subfig}
\usepackage[colorlinks=true,citecolor=blue,linkcolor=blue]{hyperref}

\begin{document}

\title{Spectroscopically Confirmed Lyman-Alpha Emitters from Redshift 5 to 7 Behind Ten Galaxy Cluster Lenses}
\author{S. Fuller$^1$, B. C. Lemaux$^1$, M. Bradač$^1$, A. Hoag$^1$, K. B. Schmidt$^2$, K. Huang$^1$, V. Strait$^1$, C. Mason$^{3,\dagger}$, T. Treu$^4$, L. Pentericci$^{5}$, M. Trenti$^{6,7}$, A. Henry$^8$, M. Malkan$^4$}
\affil{$^1$Department of Physics, University of California, Davis, 1 Shields Ave, Davis, CA 95616, USA}
\affil{$^2$Leibniz-Institut f{\"u}r Astrophysik Potsdam (AIP), An der Sternwarte 16, 14482 Potsdam, Germany}
\affil{$^3$Harvard-Smithsonian Center for Astrophysics, 60 Garden St, Cambridge, MA, 02138, USA}
\affil{$^4$Department of Physics and Astronomy, University of California, Los Angeles, CA 90095-1547, USA}
\affil{$^5$INAF Osservatorio Astronomico di Roma, Via Frascati 33, I-00040 Monteporzio (RM), Italy}
\affil{$^6$School of Physics, University of Melbourne, Parkville 3010, VIC, Australia}
\affil{$^7$ARC Centre of Excellence for All Sky Astrophysics in 3 Dimensions (ASTRO 3D), Australia}
\affil{$^8$Space Telescope Science Institute, 3700 San Martin Drive, Baltimore, MD, 21218, USA}
\affil{$^\dagger$Hubble Fellow}

\begin{abstract}
We present 36 spectroscopically confirmed intrinsically UV-faint Lyman-alpha emitting galaxies (LAEs) from follow-up observations with Keck/DEIMOS of high-redshift candidates behind 10 galaxy cluster lenses. High-redshift candidates were selected to be between $5\lesssim z \lesssim 7$ from a variety of photometric data using \textit{HST} and \textit{Spitzer} imaging surveys. We used photometric redshift information derived from accompanying photometric data to perform an integrated photometric redshift probability, or, P($z$) cut $>1\%$ between $5<z<7$ in order to construct a sample of 198 high-redshift objects, 136 primary targets and 62 secondaries (serendipitous objects). Our high-redshift sample spans intrinsic UV luminosities from a few $L^*$ down to $0.001L^*$. We identified 19 high-confidence detections of Ly$\alpha$ in our final sample and an additional 17 likely detections. Five of these detections have been previously reported spectroscopically. We find overall the emission lines to be redward skewed. We divided our sample into a lower-redshift ($z\sim5.5$) and higher-redshift ($z\sim6.5$) sample and ran Monte Carlo trials, incorporating the strength of the Ly$\alpha$ emission, the photometric redshift of the non-detections, and different treatments of multiple images. Considering only objects where Ly$\alpha$ could be detected at EW(Ly$\alpha$)$>$25Å at $3\sigma$ at the fiducial depth of our survey, and considering only those galaxies with EW(Ly$\alpha$)$>$25Å as true LAEs, and finally, considering only objects with $m_{AB}<26.8$, we found the LAE fraction to be flat, or modestly increase from 0.26$\pm0.04$ to 0.30$\pm0.04$ over the same redshift interval. These values relative to those at lower-redshift samples are consistent with the rising LAE fraction with redshift out to $z\sim6$ found in the literature, but at $z\sim6.5$ there is some tension between our results and results from surveys at intrinsically brighter luminosities. From our analyses we conclude intrinsically fainter galaxies have Ly$\alpha$ emission, and there is a steep drop in the LAE fraction from our high-redshift sample at $z\sim6.5$ and from similar galaxies at $z\sim7.5$. This likely indicates we are witnessing the tail end of the epoch of reionization, as such a drop is not expected due to changes of intrinsic galaxy properties between these redshifts.
\end{abstract}

\section{Introduction}
During the epoch of reionization (EoR), the universe underwent a phase transition in which ultraviolet (UV) photons from the first sources ionized the neutral hydrogen in the Intergalactic Medium (IGM) that dominated during earlier epochs known as the Dark Ages. Our picture of the EoR remains incomplete, as fewer than two decades have passed since the first observations of quasars beyond $z=6$ \citep{becker2001evidence,djorgovski2001threshold}. More recently, studies of the Cosmic Microwave Background with the Wilkinson Microwave Anisotropy Probe (WMAP, \citealt{bennett2013nine}) and Planck \citep{akrami2018planck} have constrained the redshift of ``instantaneous" reionization. Results from WMAP and Planck found $z_{\textrm{reion}}\sim10.6^{+1.1}_{-1.1}$ \citep{hinshaw2013nine} and $z_{\textrm{reion}}\sim8.8^{+1.7}_{-1.4}$ \citep{ade2016planck} respectively. From the perspective of probing the EoR through galaxy populations, in the past two decades, we have been able to push observations to fainter objects. Observations of galaxy and quasar spectra indicate reonization was patchy \citep{djorgovski2001threshold,pentericci2014new,tilvi2020onset}, and generally concluded by $z\sim6$ \citep{fan2006constraining}. However, it is difficult to get a complete picture of reionization by observing quasars and bright galaxies because they are rare and therefore probe a few lines of sight. Additionally, these objects are the most massive and brightest at these redshifts, and therefore do not represent the general galaxy population nor the average environments in the universe at these epochs. Intrinsically low luminosity galaxies are far more common than quasars and, with sufficient integration time, allow surveys of many lines of sight and less-biased environments. At the same time, these galaxies require long exposures and their Ly$\alpha$ photons can be attenuated by small amounts of neutral hydrogen. Reionization simulations \citep{madau1999radiative,choudhury2007searching,wise2014birth,qin2017dark} and studies of luminosity functions \citep{dayal2009lyman,bouwens2012lower,schmidt2014luminosity,Robertson_2015,bouwens2015uv,atek2015ultra} currently support the idea that first stars and galaxies are the primary ionizing sources as they outnumber and collectively outshine the first quasars. The debate is still ongoing though (see \citealt{madau2015cosmic,grazian2018contribution} for recent counterarguments).

To understand how reionization progressed, we can look for the Lyman-alpha (Ly$\alpha$) emission line from distant galaxies. Ly$\alpha$ is an excellent probe of neutral hydrogen because it is intrinsically the strongest recombination line and has a large cross-section \citep{peebles1993principles,Dijkstra_2014}, meaning it is easily resonantly scattered and attenuated by small amounts of neutral hydrogen and dust respectively. Although we can infer the state of the IGM from Ly$\alpha$ emission, making inferences on the Interstellar Medium (ISM) is difficult, as Ly$\alpha$ can have many interactions before escaping the galaxy. Currently we do not have a good understanding of the Ly$\alpha$ escape fraction ($f^{\textrm{Ly}\alpha}_{\textrm{esc}}$) at reionization redshifts, but we do know that it generally increases out to $z\sim6$ \citep{hayes2011redshift}. For a thorough discussion of Ly$\alpha$ escape in high-redshift galaxies, including the effect of galactic outflows, see e.g. \citealt{smith2018physics}, \citealt{dijkstra2017saas}.

Star formation is the main source of Ly$\alpha$ emission. Dust in the ISM can destroy the Ly$\alpha$ photons, and surrounding neutral hydrogen can scatter the Ly$\alpha$ photons. The number of photons emitted and destroyed should evolve with redshift as the star formation rate and dust content of the galaxy changes, however, given only $\sim$300 Myr passed between $z=6$ and $z=8$, the properties of the ISM of galaxies at fixed ultraviolet (UV) luminosity or stellar mass likely did not change significantly. This is supported by the lack of change in the UV continuum slopes between these redshifts (e.g. \citealt{bouwens2015uv}). Thus, a change in the number of Lyman Alpha Emitters (LAEs) between these two redshifts is likely caused by changes in the number of neutral hydrogen scattering events in the circumgalactic medium (CGM) and IGM, rather than changes in the production or destruction of Ly$\alpha$ photons inside the galaxy.

Much work has been done recently on determining the fraction of Lyman Break Galaxies (LBGs) that emit Ly$\alpha$. This test is commonly referred to as the LAE fraction test and has been performed by a variety of groups at varying redshifts (e.g., \citealt{ota2008reionization,fontana2010lack,stark2010keck,stark2011keck,pentericci2011spectroscopic,schenker2011keck,ono2011spectroscopic,mallery2012lyalpha,shibuya2012first,treu2013changing,caruana2014spectroscopy,pentericci2014new,tilvi2014rapid,oesch2015spectroscopic,zitrin2015lyalpha,furusawa2016new,de2017vlt,hu2017first,caruana2017muse,pentericci2018candelsz7}). A more complete understanding of the state of the IGM can be gained by tracking the evolution of the Ly$\alpha$ equivalent width with redshift. Such an analysis involves radiative transfer simulations to determine the evolution of neutral hydrogen with redshift through the volume averaged neutral hydrogen fraction using Ly$\alpha$ (e.g. \citealt{mason2018universe,hoag2019constraining,mason2019inferences}).

Observing LAEs, even bright ones, at reionization redshifts typically require a full night of integration on 10m class telescopes. In recent years, gravitational lensing has been used to push to higher redshifts and lower intrinsic luminosities, albeit by probing smaller effective fields of view. Currently, we do not have a complete understanding of galaxy properties at these redshifts. The James Webb Space Telescope (JWST) could followup sub $L^*$ (where $L^*$ is the characteristic luminosity of galaxies in our redshift range) galaxies to study galaxy properties, potentially at the onset of reionization. 

In this paper, we present our spectroscopic follow-up of 198 high-redshift candidates with intrinsic UV luminosities ranging from $L^*$ down to $0.001L^*$, leading to the confirmation of 36 Ly$\alpha$ emitting galaxies between redshift 5 and 7 and an analysis of the evolution of the LAE fraction over this $\sim$500Myr window in cosmic time.

In Section 2 we discuss the imaging data and photometric catalogs. We then discuss our DEIMOS target selection process, spectroscopic observations, data reduction methods, and then our photometric post-selection process to generate our candidate high-redshift sample. In Section 3 we present our search methods and analysis of the emission lines in the spectral data. In Section 4 we present the results of our search and analysis. In Section 5 we present our conclusions, including a discussion of what our results indicate about reionization and anticipate future work studying the properties of the LAEs. We adopt a $\Lambda$CDM cosmology with $H_0 = 70$ km s$^{-1}$, $\Omega_m = 0.27$, and $\Omega_{\Lambda} = 0.73$. All equivalent widths were converted to rest-frame. All magnitudes are in the AB system.

\section{Data}
In this section we discuss the imaging data, photometric catalogs, target selection process, lens models, spectroscopic observations and data reduction process, and our high-redshift sample selection process.

\begin{table*}[ht]
    \begin{center}
        \caption{Galaxy Clusters}
        \begin{tabular}{| c c c c c c c |} \hline \hline
            Cluster Name & Short Name& $\alpha_{J2000}$ & $\delta_{J2000}$ & $z_{cluster}$ & HST Imaging (\emph{Spitzer Imaging}) & HST Spectroscopy\\ \hline
            Abell 2744 & A2744 &00:14:19.51 &$-$30:23:19.18 &0.308 &HFF &GLASS\\
            Abell 370 & A370&02:39:50.50 &$-$01:35:08. &0.375 &HFF &GLASS\\
            MACSJ0416.1$-$2403 & MACS0416&04:16:09.39 &$-$24:04:03.9 &0.396 &HFF/CLASH& GLASS\\
            MACSJ0717.5+3745 & MACS0717& 07:17:31.65 &+37:45:18.5 &0.548 &HFF/CLASH (\emph{SURFSUP})& GLASS\\
            MACSJ0744.8+3927 & MACS0744& 07:44:52.80& +39:27:24.4&0.686 &CLASH (\emph{SURFSUP})& GLASS\\
            MACSJ1149.5+2223 & MACS1149& 11:49:35.86 &+22:23:55.0 &0.544 &HFF/CLASH (\emph{SURFSUP})& GLASS\\
            MACSJ1423.8+2404 & MACS1423& 14:23:47.76 &+24:04:40.5 &0.545 &CLASH (\emph{SURFSUP})& GLASS\\
            MACSJ2129.4$-$0741 & MACS2129& 21:29:26.06&$-$07:41:28.8 &0.570 &CLASH (\emph{SURFSUP})& GLASS\\
            MACSJ2214.9$-$1359 & MACS2214& 22:14:57.41  &$-$14:00:10.8 &0.500&SURFSUP (\emph{SURFSUP})&-\\
            RXJ1347.5$-$1145 & RXJ1347& 13:47:30.59 &$-$11:45:10.1 &0.451 &CLASH (\emph{SURFSUP})& GLASS\\
            \hline
        \end{tabular}
    \end{center}
All GLASS clusters have a photometric preselection of the high-redshift spectroscopic sample \citep{schmidt2016grism} described in Section 2.4.
\end{table*}

\subsection{Imaging Data}
Our candidate high-redshift objects are gravitationally lensed by 10 galaxy clusters that were selected from a variety of overlapping \textit{HST} and \textit{Spitzer} programs. Five clusters are Hubble Frontier Fields (HFF, \citealt{lotz2017frontier}, the sixth cluster is not visible from Keck), with a 5$\sigma$ limiting AB magnitude of $\sim$29 in all filters (see below for details on which filters were observed). Four clusters are part of the Cluster Lensing And Supernova Survey with Hubble (CLASH, \url{http://www.stsci.edu/~postman/CLASH/}), \citealt{postman2012cluster}), a large \textit{HST} imaging program that identified and characterized galaxies at $z > 7$ behind 25 galaxy clusters with a $5\sigma$ limiting AB magnitude in F160W of 27.7. The last cluster is from the Spitzer UltRa Faint SUrvey Program (SURFSUP, \citealt{bradavc2014spitzer}), a joint \textit{Spitzer} and \textit{HST} imaging program to study intrinsically faint gravitationally-lensed galaxies at $z>7$. Of our 10 clusters, nine have spectroscopic data from the Grism Lens-Amplified Survey from Space (GLASS, \citealt{schmidt2014through,treu2015grism}), a spectroscopic \textit{HST} program that analyzed the light from gravitationally-lensed background galaxies. 

The HFF clusters were observed in seven filters; F435W, F606W, and F814W from the Advanced Camera for Surveys (ACS, \citealt{sirianni2005photometric}) and F105W, F125W, F140W, and F160W from the Wide Field Camera 3 (WFC3, \citealt{kimble2008wide}). The CLASH clusters were observed in the same filters as the HFF clusters, in addition to the following ACS and WFC3 filters: F435W, F475W, F555W, F625W, F775W, F850LP, and F110W. One SURFSUP cluster was observed from \textit{HST} in F105W, F125W, and F160W.

Table 1 summarizes the basic properties of our 10 galaxy clusters and their corresponding \textit{HST} and \textit{Spitzer} programs. These specific clusters were chosen for their excellent magnifying properties. In general these fields had the greatest number of high-redshift candidates of the clusters in the surveys listed above. See Section 2.6 for the high-redshift candidate selection process.

\subsection{Photometry}
A photometric pipeline was developed by the ASTRODEEP (\url{http://www.astrodeep.eu/}) team that was used to measure photometry for the six HFF clusters: A2744, MACS0416 \citep{merlin2016astrodeep,castellano2016astrodeep}, MACS0717, MACS1149 \citep{di2017astrodeep}, RXJ2248, and A370 \citep{bradac2019hubble}, of which we study five here. For our remaining five clusters, an identical method to the ASTRODEEP method was used: MACS2129, RXJ1347, MACS1423 \citep{huang2016spitzer}, MACS0744, and MACS2214 \citep{hoag2019constraining}. Briefly, the process involved generating point-spread function (PSF) matched \textit{HST} images using {\tt{psfmatch}} in {\tt{IRAF}}, making sure to PSF match all of the \textit{HST} images with the F160W images. In order to improve the detection of faint objects, intracluster light (ICL) was subtracted using {\tt{\textsc{Galfit}}}. However, the ICL subtractions were not performed for MACS0744 and MACS2214 because the high-redshift objects in these clusters are not heavily contaminated by the ICL, due in part to the shallower \textit{HST} images in these clusters. \emph{HST} photometry was then measured using {\tt{SExtractor}} \citep{bertin1996sextractor} on all images using the stacked \textit{HST} WFC3 near-infrared images as a detection image. For the sake of brevity, we will refer to these catalogs collectively as the ASTRODEEP catalogs. From the ASTRODEEP catalogs, distributions of photometric redshift probabilities, P($z$)s, were generated by fitting the photometry using EAzY \citep{brammer2008eazy}, using a flat prior on $z$ (because all of our galaxies are gravitationally lensed) and isophotal magnitudes.

For those objects with \emph{Spitzer} images, photometry was measured using {\tt{T-PHOT}} \citep{merlin2015t}, software developed by the ASTRODEEP team, designed to extract fluxes in crowded fields with different point spread functions.

In addition to the photometric catalogs mentioned above, we also had access to the CLASH catalogs \citep{molino2017clash}. Information from these catalogs was used for 18 objects that were unavailable in the ASTRODEEP catalog. For CLASH objects, we had access to the most likely redshift, defined as the peak of the P($z$), and the upper and lower 2$\sigma$ values. See Appendix A for a comparison between the ASTRODEEP and CLASH photometric catalogs. We describe how we use this information to generate our high-redshift sample in Section 2.5.

\subsection{Lens Models}
Our candidate high-redshift objects are gravitationally lensed and therefore require us to have a good understanding of the lensing properties of the cluster to calculate their intrinsic luminosities. The lens models used to generate magnification information were developed by \citet{bradavc2005strong,bradavc2009focusing}. We selected these models because they exist uniformly across all of the clusters. The model reconstructs the gravitational potential using an adaptive pixel grid, allowing for a solution across a larger FoV and increased resolution near the cluster center and in the vicinity of multiple images. A $\chi^2$ fit is done between the data, comprised of the positions of multiple images from strong lensing and the ellipticity of galaxies from weak lensing, and their lens model-predicted values. From the best-fit potential we find the magnification for our high-redshift candidates using the photometric redshift, position, and the magnification map at the appropriate redshift from the lens model. Six of our objects had a peak photometric redshift less than the redshift of the cluster, meaning lensing would not be possible if the photometric redshift was trusted. However, these objects had secondary peaks in their P($z$)s within our redshift window, so we decided to set their redshift to 6 when calculating the magnification as magnifications do not vary significantly above $z\sim5$. Six objects were outside the lens model FoV. For these objects we set the magnification to one. With increasing distance from the cluster center, the magnification should approach one, and the DEIMOS footprint extends past the virial radii of our clusters, so the approximation is sufficient.
 
\subsection{DEIMOS Target Selection}
We used the DEep Imaging Multi-Object Spectrograph (DEIMOS, \citealt{faber2003deimos}) on Keck for our followup observations of 198 promising high-redshift candidates. In our fiducial setup (27 of our 32 masks), DEIMOS operates from $\sim$7500Å to $\sim$10000Å with a 1200G grating and central wavelength of 8800Å. The Ly$\alpha$ emission line from objects between $5.2\lesssim z \lesssim 7.2$ falls within this wavelength range. For a more detailed description of our DEIMOS setups, see Section 2.5.


136 of our 198 high-redshift candidates were DEIMOS targets, selected via a variety of heterogeneous criteria from the photometry, including color and dropout selections, and photo-$z$ cuts, and spectroscopic data available at the time, including potential GLASS detections. The majority of these objects came from a photometric preselection \citep{schmidt2016grism}. About 28\% of the objects from the pre-selection made it onto our masks. The remaining 62 were serendipitous, or, secondary objects, in which the slit happened to fall across the photometrically preselected object, and 50\% of the object's rest-frame UV centroid was visually estimated to be within the slit boundary. In general, the masks were designed to maximize the number of high-redshift objects we could observe. We assigned a priority to objects based on the peak of their P($z$) from the best photometric catalog available at the time, as the final data was sometimes not available at the time. Note, this catalog was not necessarily the same catalog we used for our analysis. Objects with peaks at $z>6$ would be given a high priority and objects with peaks between $5<z<6$ would be given a lower priority, yet still higher than priorities for objects in filler slits. The P($z$) information contained information from \textit{Spitzer} observations when available at the time of making the mask. Objects with spectroscopic detections in GLASS were assigned higher priorities than those without GLASS detections. The effects of the GLASS selection are described in Section 4.2. The slitmasks were designed using DSIMULATOR. In the remaining regions inside the \textit{HST} Field of View (FoV), masks were filled with arcs and lower redshift objects, and in regions outside \textit{HST} FoV, with potential cluster members or other lower-priority objects.

\subsection{Spectral Observations and Data Reduction}
We observed with DEIMOS over the course of nearly four and a half years. The first observation occurred on April 3$^{\textrm{rd}}$, 2013 UTC and the most recent occurred on September 28$^{\textrm{th}}$, 2017 UTC. Table 2 lists our 32 DEIMOS masks, exposure times, and observing conditions. Two masks observed in 2013 used the 600ZD grating with a central wavelength of 8000Å, DEIMOS coverage from $\sim$5000Å to $\sim$10000Å, and a pixel scale of $0.66$Å/px. Three masks observed in 2013 used the 830G grating with a central wavelength of $\sim$8100Å, DEIMOS coverage from $\sim$6300Å to $\sim$10000, and a pixel scale of $0.47$Å/px. All other masks used the 1200G grating with a central wavelength of 8800Å and a pixel scale of $0.33$Å/px. The total exposure time for each mask ranged from about one hour up to 10 hours. High-redshift candidates typically appeared on multiple masks. The average exposure time per mask was about three hours. Our median $1\sigma$ flux depth over all of our observations was 1.6$^{+1.0}_{-0.7}\times10^{-18}$ erg/s/cm$^2$, where the uncertainties represent the 16$^{\textrm{th}}$ and 84$^{\textrm{th}}$ percentiles. Typical seeing was $\lesssim$ 1$^{\prime\prime}$ and attenuation $\lesssim$ 0.1 mag. Average seeing was determined by looking at the 1D profile of a fine alignment star for each exposure of a mask and calculating its exposure time-weighted Full Width at Half Maximum (FWHM). Sky attenuation values (if available) were measured by SkyProbe (\url{https://www.cfht.hawaii.edu/Instruments/Skyprobe/}). These values are estimates and are presented in Table 2 as an overview. They are not used in flux calibration.

The data were reduced with a modified version of the DEIMOS \textsc{spec2d} pipeline described in \citet{cooper2012spec2d} and \citet{newman2013deep2}. First, calibration files were generated from the flatfield frames (with pixel-to-pixel variation and cosmic ray corrections) and the arc frames. Next, a wavelength solution was generated and then flat-fielded and rectified using the calibration files. Then an inverse variance image was produced. Next, a sky-subtraction was performed using a model sky spectrum, and a cosmic ray cleaning was done. Finally, all of the individual science exposures were combined using inverse variance weighting. The inverse variance array from the pipeline was used for our noise analysis. For more details on the reduction process see \citet{newman2013deep2}, as well as \citet{lemaux2019persistence} for specific changes made in our version of the pipeline. Certain exposures with very short integration times, large seeing, or high sky attenuation values were excluded from the reduction process to maximize the resultant signal-to-noise ratio (S/N). From our reduced 2D data, one dimensional spectra were extracted using a boxcar extraction at the cataloged positions of the targeted objects used to generate the DEIMOS masks. The boxcar extraction was preferred over the Horne extraction \citep{horne1986optimal} because the overall S/N of all candidates when collapsed over the full wavelength range is very low as the continuum is generally completely undetected. For secondary high-redshift objects, the 1D spectra were re-extracted at the measured position of the high-redshift object.

\begin{table*}[ht]
\begin{center}
\begin{threeparttable}
\caption{Mask and Observing Information}
\begin{tabular}{| c c c c c c c c |} \hline
Cluster & Mask& Grating & \parbox{2.5cm}{\centering Observation Date\\ (UTC)}&\parbox{0.5cm}{\centering $t_{\textrm{exp}}$\\ (sec)}&\parbox{2cm}{\centering $<$Seeing$>$\tnote{1}\\ ($^{\prime\prime}$)}&\parbox{2cm}{\centering Attenuation\tnote{2}\\ (mag.)}&Airmass\\\hline \hline
A2744&274415B1&1200G&2015oct16&3600&1.05&-&1.62-1.75\\
A2744&274415B1&1200G&2015nov17&4620&0.78&0.03-0.09&1.56-1.61\\\hline
A370&A370\_D3n&1200G&2014sep01&3600&0.54&0-0.01&1.13-1.23\\
A370&A370\_D4n&1200G&2014sep01&5400&0.59&0-0.03&1.08-1.10\\
A370&A37017B1&1200G&2017sep28&8400&0.81&0.04-0.07&1.07-1.23\\
A370&A37017B2&1200G&2017sep28&7200&0.97&0.04-0.06&1.09-1.33\\\hline
MACS0416&041615B1&1200G&2015oct16&3756&0.99&-&1.38-1.44\\
MACS0416&041615B1&1200G&2015nov13&7200&1.38&0.2-1.4&1.40-1.53\\
MACS0416&041615B1&1200G&2015nov17&11615&0.80&0.02-0.08&1.38-1.81\\\hline
MACS0717&071715B1&1200G&2016jan06&12000&0.81&0.03-0.2&1.05-1.48\\\hline
MACS0744&M744D\_1&1200G&2014nov28&6967&0.65&0.06-0.25&1.10-1.35\\
MACS0744&M744D\_1&1200G&2014nov29&6000&0.87&-&1.06-1.16\\
MACS0744&M744D\_2&1200G&2014nov28&9600&0.67&0.05-0.26&1.06-1.16\\
MACS0744&074416A1&1200G&2016feb06&12000&0.95&0-0.04&1.06-1.15\\
MACS0744&074416A1&1200G&2016feb07&8400&1.05&0.02-0.08&1.14-1.64\\
MACS0744&074416A1&1200G&2016mar11&15900&1.05&0.01-0.08&1.06-1.55\\
MACS0744&074416A2&1200G&2016feb06&8400&0.79&0.01-0.08&1.07-1.35\\
MACS0744&074416A2&1200G&2016feb07&2400&1.08&0.04-0.08&1.06-1.11\\\hline
MACS1149&miki11D&830G&2013apr04&19200&0.62&0.08-0.13&1.00-1.49\\
MACS1149&114915A1&1200G&2016jan06&3480&0.58&0.05-0.15&1.00-1.02\\
MACS1149&114916A1&1200G&2016feb07&7200&1.46&-&1.08-1.46\\\hline
MACS1423&miki14D&600ZD&2013apr03&10800&0.70&0.08-0.14&1.00-1.31\\
MACS1423&miki14D2&600ZD&2013apr04&3000&1.01&0.08-0.12&1.22-1.41\\
MACS1423&miki14D2&600ZD&2013apr05&6000&1.42&0.2-1.1&1.08-1.34\\
MACS1423&C14215A1&1200G&2015may14&8400&0.86&-&1.00-1.09\\
MACS1423&C14215A1&1200G&2016feb06&10800&0.73&0.03-0.1&1.00-1.36\\
MACS1423&C14215A2&1200G&2015may15&9600&1.40&0.04-0.12&1.02-1.42\\
MACS1423&C14215A2&1200G&2015may16&3600&1.32&-&1.27-1.49\\
MACS1423&C14215A2&1200G&2016feb07&8973&1.94&-&1.00-1.52\\
MACS1423&142317A1&1200G&2017may24&4200&0.76&0.07-0.12&1.15-1.34\\
MACS1423&142317A1&1200G&2017may25&12600&0.71&0.07-0.11&1.00-1.24\\
MACS1423&142317A2&1200G&2017may25&6900&0.59&0.07-0.12&1.07-1.37\\\hline
MACS2129&M2129\_D3&1200G&2014sep01&10800&0.58&0-0.04&1.15-1.94\\
MACS2129&M2129\_D4&1200G&2014sep01&10800&0.79&0-0.04&1.13-1.42\\
MACS2129&212915A1&1200G&2015may15&4800&1.09&0.02-0.12&1.30-1.65\\
MACS2129&212915A1&1200G&2015may16&6000&0.99&0.02-0.09&1.23-1.65\\
MACS2129&212915B1&1200G&2015oct16&7800&1.18&0-0.06&1.13-1.21\\
MACS2129&212917A1&1200G&2017may24&4600&0.94&0.09-0.11&1.23-1.43\\\hline
MACS2214&221417B1&1200G&2017sep28&8400&1.07&0-0.07&1.24-1.80\\
MACS2214&221417B2&1200G&2017sep28&8400&0.81&0.04-0.06&1.20-1.35\\\hline
RXJ1347&miki13D&830G&2013apr04&6000&0.74&0.08-0.11&1.26-1.50\\
RXJ1347&miki13DG&1200G&2014may27&7158&0.91&0.06-0.17&1.18-1.38\\
RXJ1347&miki13DG2&1200G&2014may27&6000&0.90&0.06-0.13&1.19-1.39\\
RXJ1347&134717A1&1200G&2017may24&14400&0.84&0.06-0.12&1.18-1.36\\\hline
\end{tabular}
\newpage
\begin{tablenotes}

\item[1]To calculate the average seeing for a given mask and date, we fit a Gaussian to the profile of a bright alignment star and calculated exposure time-weighted seeing values for each frame. 
\item[2]Attenuation values were estimated from SkyProbe. Attenuation was unavailable for dates marked "-". This value is for reference only and is not used in our formal flux calibration process.
\end{tablenotes}
\end{threeparttable}
\end{center}
\end{table*}

\subsection{High-Redshift Candidate Sample}
We refined our initial high-redshift sample to create a more coherent collection of high-redshift candidates. We matched our initial high-redshift candidates with the corresponding P($z$) information in the ASTRODEEP catalog (if available) and calculated the integrated P($z$) between $z=5$ and $z=7$ for each object. For the 18 objects not in the ASTRODEEP catalog (outside the WFC3 filter FoV), we used information from the CLASH catalog. For the CLASH objects, we did not have access to the full P(z) information, so we crudely estimated the integrated P($z$) using the given photo-$z$, and the upper and lower 2$\sigma$ values. We included any object with an integrated P($z$) $>1$\% within the target redshift range in our LAE/LBG analysis. This cut is very inclusive, but in our LAE fraction analysis (Sections 4.1 and 4.2), objects with a high probability of being at lower redshift are severely down-weighted. The same $>1$\% cut was used by \citet{hoag2019constraining}, our parallel MOSFIRE campaign targeting $z\sim7.5$ galaxy candidates in the GLASS clusters. Additionally, we removed all objects that were fewer than 10 pixels ($\sim1.2^{\prime\prime}$) away from the edge of the slit along the spatial dimension. Poor sky-subtraction along the slit edge, and fringing at the extreme red end of our DEIMOS spectral window can cause artifacts that can be mistaken for emission lines. This left us with the 198 high-redshift candidates. 136 of these candidates were targets of DEIMOS slits and the remaining 62 were secondaries. In this paper we present the detections and calculate the LAE fraction for the final high-redshift candidate sample. The non-detections will be presented as part of an analysis in Lemaux et al. in prep.

In Figure 1 we plot the apparent magnitudes in F160W ($m_{\textrm{F160W}}$) for our high-redshift sample, separated by targets and secondaries. Figure 2 shows the P($z$) information for the 180 objects for which we have full P($z$)s. The 18 objects with crude P($z$)s reconstructions are excluded from the plot. Also plotted are composite P($z$)s, again separated by targets and secondary objects. The composite P($z$) was calculated by combining all of the normalized individual P($z$)s. It is normalized such that the total area under the curve for the complete sample has probability of unity.

\begin{figure}[H]
  \includegraphics[width=\linewidth]{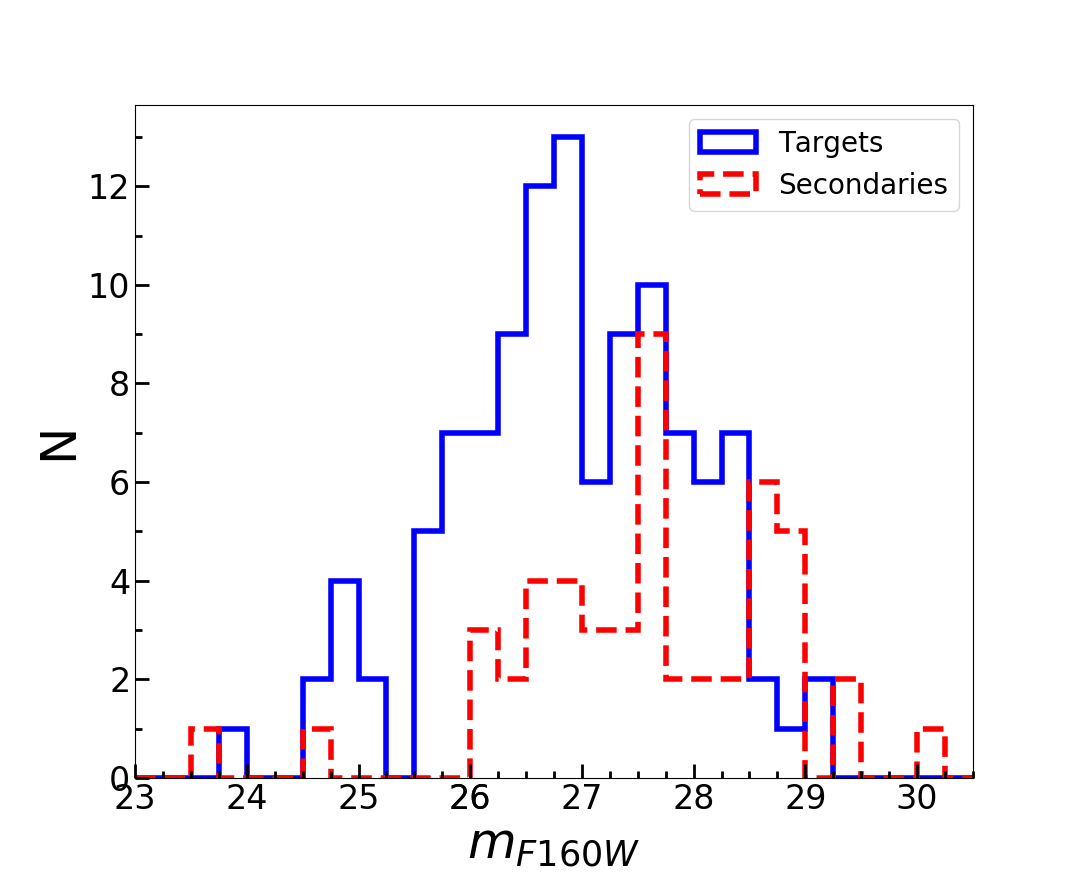}
  \caption[labelsep=<colon>]{Histogram of F160W apparent magnitudes, $m_{\textrm{F160W}}$, for our sample of 198 high-redshift objects.}
\end{figure}

\begin{figure}[H]
  \includegraphics[width=\linewidth]{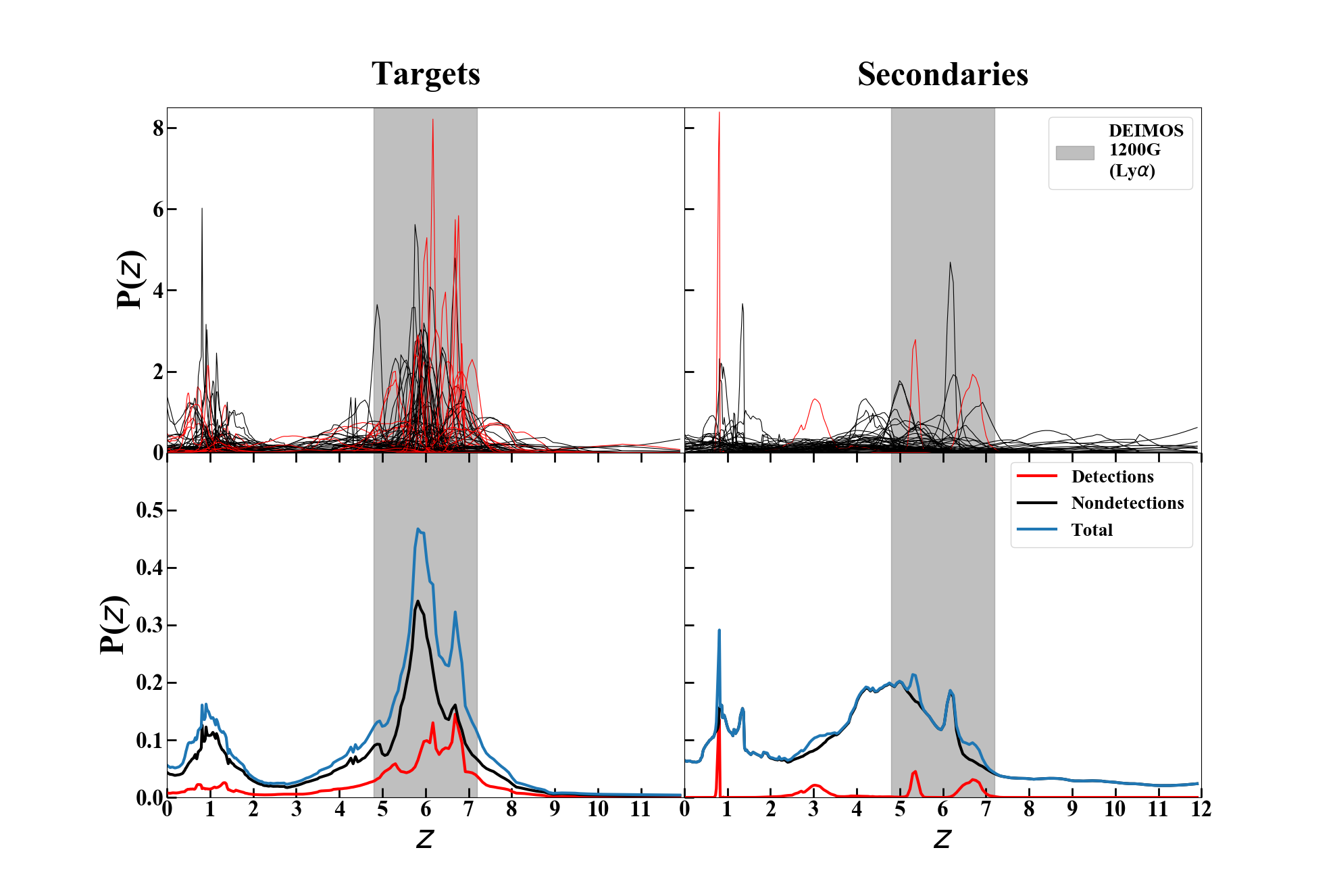}
  \caption{Top: Individual P($z$)s of 180 of our 198 DEIMOS-observed high-redshift objects for both targets (left) and secondary objects (right). The remaining 18 have only crude P($z$) reconstructions from CLASH and are excluded from the plot. Bottom: Composite P($z$)s of targets (left) and secondary objects (right). The entire sample is in blue, the detections are in red, and the nondetections are in black. The grey area indicates the DEIMOS coverage for Ly$\alpha$ for the 1200 line grating (used for 27 of our 32 masks) with a central wavelength of 8800Å.}
\end{figure}

\section{Analysis}
Here we describe our visual and automated search methods for Ly$\alpha$ emission lines and present our luminosity and equivalent width (EW) calculations for our detections.

\subsection{Visual Search}
We did a preliminary search by eye for Ly$\alpha$ emission lines for all objects in our sample by first stacking the corresponding 1D and 2D spectra using inverse-variance weighting for all dates and masks that the object was observed on. If multiple gratings were used for a given object, the wavelength array was interpolated onto a standard grid with a resolution of the poorest resolution grating. We then inspected the stacked spectra across the entire DEIMOS wavelength coverage. We looked at the smoothed 2D spectra using different Gaussian smoothing kernels to increase the contrast of any diffuse emission lines. Possible emission lines at or near the spatial location of the object were noted and then checked in each of the individual nights for confirmation when available. The visual search was performed systematically by one person, however, a few detections that were noted from previous informal searches were verified. We followed up our visual search with a formal signal-to-noise ratio search.

\subsection{Signal-to-Noise Ratio Search}
We performed an automated integrated signal-to-noise ratio search of the stack of the boxcar-extracted 1D spectra. The search identified all locations where the integrated signal-to-noise ratio within a wavelength search width of 10Å was above three. The search output a list of all the detections with their central wavelength, a 1D spectrum indicating the locations, and 2D cutouts at all the locations. The search excluded regions closer than 2.5Å from the edge of a skyline on either side, however, we decreased this for emission lines found by-eye that were on top of skylines. This means we may have missed faint emission lines that were coincident with skylines that were not found in the visual search described previously. The skyline information was taken from a random slit from a typical mask utilizing each of the three gratings that were used. For 600ZD, 830G, and 1200G we included all skylines with pixel counts above 300, 150, and 100 respectively. These cuts included $\sim$80\% of skylines of all strengths, and therefore should be sufficient for the search algorithm. Additionally, options were available to change the 1D and 2D smoothing kernel size, the search window in which the integrated S/N is calculated, and the minimum distance from a skyline that would be searched. All of the detections from the automated search were visually checked in the 2D spectrum. If a by-eye detection was identified in the stacked 2D spectrum as well, we made additional checks to convince ourselves the emission line was Ly$\alpha$.

We visually inspected the \textit{HST} image and checked our photometric catalogs to see if there were any foreground objects with a close angular separation that could contaminate the light from the high-redshift object. We looked at the extent of the emission line in the spatial dimension to rule out cosmic rays and artifacts from the skyline subtraction. We have noted some physical Ly$\alpha$ offsets in our sample (Lemaux et. al in prep.), as has been seen at some level at lower redshifts (e.g., \citealt{hoag2019constrainingoffset}).

Although we didn't detect strong continuum for any of our candidates, we could have expected to see other emission lines if the candidate was a low-redshift interloper. A common line that is seen among low-$z$ galaxies in DEIMOS is H$\alpha$. This line would fall in the DEIMOS window for galaxies at 0.15$<z<0.5$. Additionally, we have the spectral resolution using the 1200G grating to be able to distinguish the [OII] doublet at $\lambda$3726Å and $\lambda$3729Å for objects between $1.0<z<1.7$ and the spectral baseline to identify two [OIII] lines at $\lambda$4959Å and $\lambda$5007Å for objects between $0.5<z<1$, which make up our most probable contaminants based on our composite P($z$)s in Figure 2. To improve our confidence in the Ly$\alpha$ detection, we checked for associated emission lines, such as H$\beta$ assuming an emission of [OIII], or [OIII] assuming an emission of H$\beta$. We also checked for [NII], [SII], and [OIII] assuming an emission of H$\alpha$. We were able to find a possible $z=1.063$ interloper, which we identified with a low SNR [OII] doublet with observed wavelengths of 7688Å and 7693Å. However, because the [OII] interpretation is not definitive, we decided to retain it in the high-redshift sample. None of our other detections have detectable accompanying emission lines. To further tease out contaminants, we can distinguish between typically symmetric emission lines (H$\alpha$, H$\beta$, [OIII]) or blueward-skewed lines ([OII]), and redward-skewed asymmetric ones (Ly$\alpha$). The asymmetry calculation and analysis is discussed in Section 3.3.

We assigned each of the detections that met the criteria above a quality from one to four. 
\begin{itemize}
\item A quality four (Q4) detection is a detection with $S/N>5$ that is seen across multiple observing nights, or, if there was only a single night of observation, also has a convincing shape in 2D that is spatially extended rather than point-like, or is redward skewed (see Section 3.3). These are the detections we are most confident in.
\item Quality three (Q3) was assigned to detections with $S/N\sim5$, but only have one night of observation. A quality three was also assigned to detections with $S/N>5$ in only the nights with the best observing conditions and longest exposure times.
\item Quality two (Q2) detections had a $S/N\sim5$, but were generally more point-like and less extended in the dispersion direction than quality three detections.
\item Quality one (Q1) detections had $S/N<5$ and may have only been seen in the night with the longest exposure time and best observing conditions. We are less confident in these detections, but we include them in our analysis as they were identified visually, are unlikely to be artifacts or lower-redshift lines, and in nearly all cases have a S/N$>$3. We emphasize that even though these are our lowest quality detections, they are still very likely to be high-redshift galaxies.
\end{itemize}
In Figure 10 we present 2D and 1D cutouts of all of our 36 detections along with their detection qualities. Also shown are \textit{HST} images in F160W with the slits in red and the object location within the slit in yellow. In Table 3, we present the properties of our 36 detections. 32 were detected in the ASTRODEEP catalog, and the remaining four were detected only in the CLASH catalog. Five of our detections are multiple images and are treated as individual objects in the table. A few of our detections are on or near skylines. These are marked in the tables with an asterisk. Spectral centroids and their associated errors were calculated using the non-parametric method described in \citet{teague2018robust}. This method works for both symmetric and asymmetric line profiles.

\subsection{Asymmetry}
To increase our confidence that our detections are in fact, genuinely Ly$\alpha$, we calculated the asymmetry of each emission line and compared it to what we expected for Ly$\alpha$ at reionization redshifts. Ly$\alpha$ line profiles can have a variety of shapes, in part due to complex geometries in the ISM \citep{verhamme20063d,neufeld1991escape}. Some of the line attributes include singly or doubly-peaked, symmetric or asymmetric, and Voigt or P-Cygni-like profiles. Galaxies at reionization redshifts tend to have redward skewed emission lines due to a blueward line flux suppression from neutral hydrogen \citep{dawson2007luminosity,dijkstra2016constraining}. To calculate asymmetry, first we found the peak flux density value of the stacked one dimensional spectrum. Then we moved out in the blue and red directions to locate the first instance the flux density dropped to 10\% of its peak value \citep{rhoads2004luminous}. At lower S/N, this method will give us more reliable measurements than a parametric fit. The asymmetry was then calculated as $(\lambda_R - \lambda_0)/(\lambda_0 - \lambda_B$), where $\lambda_0$ is the peak value. For redward skewed lines, asymmetry values are $>1$. The asymmetry values are quoted in Table 3. Typical asymmetry values for LAEs at $z\sim 5$ are greater than 1.3 \citep{lemaux2009serendipitous}. 12/19 of our Q4 detections and 10/17 of our Q3, Q2, and Q1 detections have an asymmetry $\geq1.3$, with averages of 2.8 and 3.0 respectively. This asymmetry is particularly apparent with our very high S/N detections.

To determine more robustly the asymmetry of our low S/N detections and understand the general profile of the Ly$\alpha$ line at these redshifts, we inverse variance-weighted mean-stacked our rest-framed 1D spectra. See Figure 3 for the profiles for our entire collection of detections, only the Q4 detections, and finally, only Q1-Q3 detections. The profiles of the Q4 detections have a distinct tail on the red end. The tail is even apparent in our Q1-Q3 detections. The asymmetry values for the all the detections, only the Q4 detections, and only the Q1-Q3 detections are 2.2, 1.8, and 2.8 respectively. The fact that we see an even larger asymmetry among the Q1-Q3 than among the Q4 detections likely indicates that many of these lower-confidence detections are in fact Ly$\alpha$. That being said, low S/N detections are more difficult to centroid well, which may be why we see a wider spectral profile among the Q1-Q3 detections.

\begin{figure}[H]
  \centering
  \includegraphics[width=\linewidth]{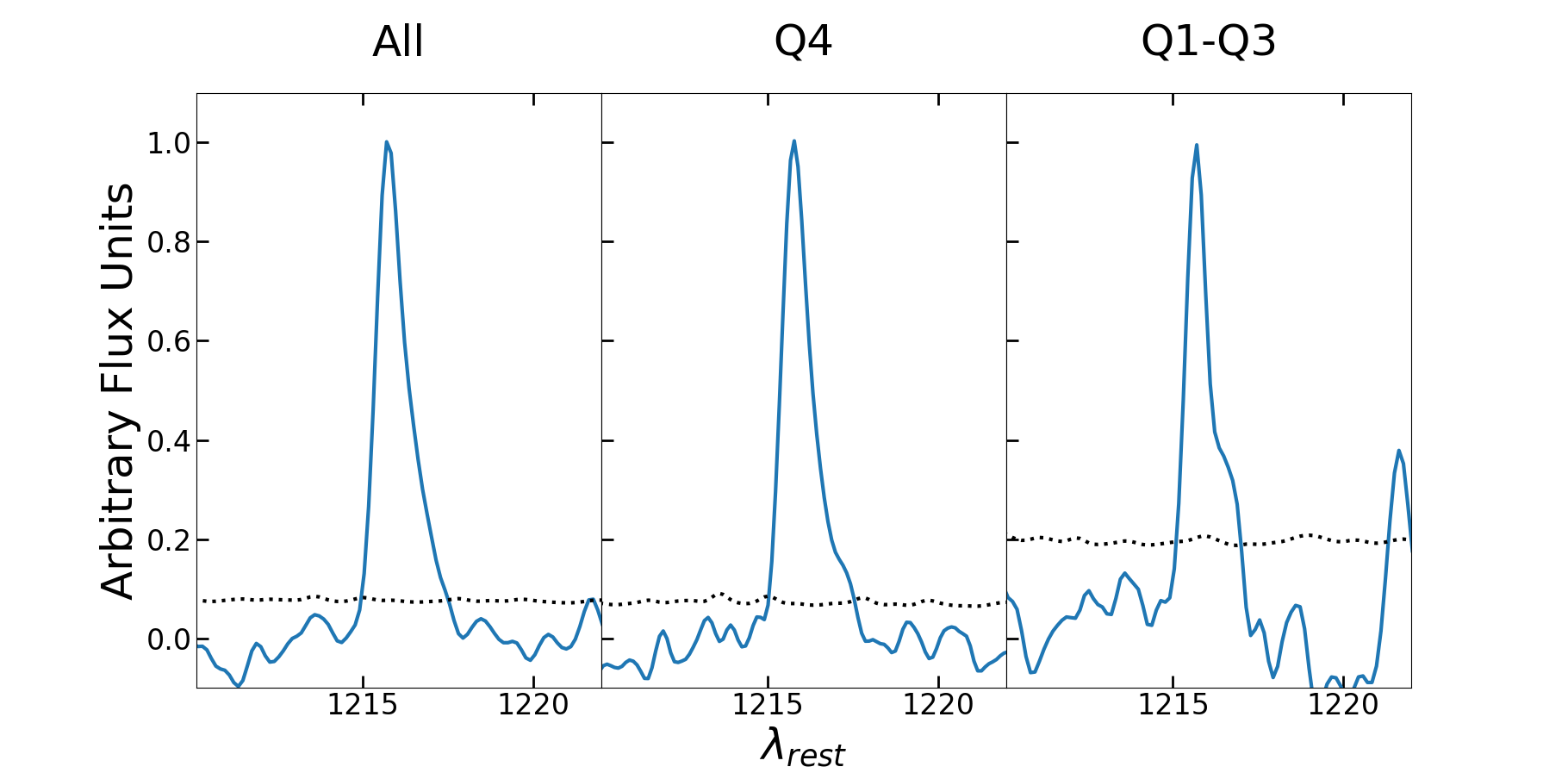}
  \caption{Stacked and smoothed (for presentation purposes) 1D rest-framed detections for all detections (left) with an asymmetry value of 2.2, only Q4 detections (center) with an asymmetry value of 1.8, and Q1-Q3 detections (right) with an asymmetry value of 2.8. Measurements were performed on the unsmoothed data. The Gaussian smoothing kernel had a standard deviation of 1.5 pixels, which corresponds to 0.50Å, 0.71Å, and 0.99Å for 1200, 830, and 600 line gratings respectively.}
\end{figure}

\subsection{Flux Calibration}
Instead of using a standard star for flux calibration, we opted to use bright objects on the mask. The benefit of this method is that the seeing and attenuation of our bright objects should be nearly identical to our high-redshift targets since we are observing them simultaneously. We selected the bright objects based on the strength of their continuum and the size and orientation of the object in relation to the slit. Ideally these objects would be in the \textit{HST} F814W footprint so we could measure the intrinsic half-light radius. For bright objects outside this footprint, we used an adjacent filter, either F775W or F850LP. Two of our bright objects were stars, and all others were foreground galaxies. For each night of observation we predicted the slit-loss for our bright objects and LAEs using simulations \citep{lemaux2009serendipitous}. This required inputting the seeing, measured from the 1D profile of a fine alignment star, the measured airmass values, the half-light radius of the object, and its position in the slit. To get the position in the slit, we measured the UV position of the object along the minor axis and take this into account when calculating the slitloss. This was performed for both secondaries and the targets. In two cases (ID 1423.38 and ID 370.55) there was a misalignment between the slit and the target, which we corrected for. Though we made this correction, the UV position could be offset from the Ly$\alpha$ position. This could be a concern for objects near the edge of the slit (ID 0717.53 and ID 0416.89). During the slit-loss correction, we also corrected for the known instrument response. The half-light radii for our bright objects were measured using {\tt{SExtractor}} on the observed-frame optical \textit{HST} image used for flux calibration. We assumed a half-light radius of 0.2$^{\prime\prime}$ for our LAEs, which is a typical radius for LAEs at reionization redshifts \citep{pentericci2018candelsz7}. At radii this small, the seeing disc dominates, which is why we decided to use a constant half-light radius. That being said, there is evidence of pervasive extended Ly$\alpha$ between $3<z<6$ \citep{steidel2011diffuse,leclercq2017muse,wisotzki2018nearly}. If the Ly$\alpha$ emission is spatially larger than the UV continuum, then we will under-predict the slit-loss. Once we predicted the slit-loss for bright objects, we calculated the actual slit-loss. First we created a spectral magnitude by convolving the 1D spectrum for each bright object with the filter transmission curve. Then we looped over different slit-loss values until the photometric magnitude of the bright object was recovered. Using the predicted slit-loss values for our bright objects and LAEs, and the actual slit-loss values for our bright objects, we calculated the actual slit-loss for our LAEs using

\begin{equation}
\frac{\Omega_{\text{Actual,LAE}}}{\Omega_{\text{Predicted,LAE}}} = \frac{\Omega_{\text{Actual,BO}}}{\Omega_{\text{Predicted,BO}}},
\end{equation} where $\Omega_{\text{BO}}$ is the slit-loss for a bright object. Four masks did not have bright objects in the \textit{HST} field-of-view. For these masks we used bright objects on other masks that were observed on the same night in which the observing conditions did not change significantly. To determine if we could reliably use the proxy masks for flux calibration, we calculated the actual / predicted slit-loss for all masks observed on the same date and found the uncertainties between all pairs of masks, 

\begin{equation}
\frac{\Omega_{i\textrm{,a/p}} - \Omega_{j\textrm{,a/p}}}{\Omega_{j\textrm{,a/p}}},
\end{equation} where $\Omega_{i\textrm{,a/p}}$ and $\Omega_{j\textrm{,a/p}}$ are the actual slit-loss values divided by the predicted slit-loss values for given masks $i$ and $j$. The mean for our uncertainties was -0.1 with 84th and 16th percentile values of 0.27 and -0.37 respectively.

Not all of the bright objects we observed had entries in the ASTRODEEP and ASTRODEEP-based photometric catalogs. For these objects, we used CLASH photometry for flux calibration.

In order to get a robust measure of the flux, we must know exactly where the object is in the sky. We checked for any astrometric shifts between the RA and Dec values used to design the DEIMOS masks and the RA and Dec values in SDSS (when available, \citealt{gunn20062,eisenstein2011sdss,alam2015eleventh}). Our average RA and Dec offsets were both 0.00$^{\prime\prime}$ with a normalized median absolute deviation (NMAD) of 0.07$^{\prime\prime}$ and 0.06$^{\prime\prime}$ for RA and Dec respectively. We also compared the DEIMOS design positions with those in the ASTRODEEP catalog. We found an average RA offset between the two of 0.00$^{\prime\prime}$ with an NMAD of 0.03$^{\prime\prime}$ and an average Dec offset of 0.02$^{\prime\prime}$ with an NMAD of 0.05$^{\prime\prime}$. We ignore this bias and estimate that our flux calibration process is accurate to better than 40\% overall, with the majority of our error coming from quantifying the slit-loss.

Line luminosities were calculated by first performing a boxcar extraction on the individual 1D spectra of high-redshift candidates coming from different masks over a spectral region encompassing the entire emission line, determined by eye. The typical extraction spanned nine or ten pixels. Additional boxcar extractions were performed blueward and redward of the emission line to estimate the background flux. The background was then subtracted from the emission region to get the line flux. To get the total line flux for each detection, we averaged the fluxes from individual nights and weighted each of the individual nights by the inverse flux error, which we calculated after flux calibrating the inverse variance array from the reduction pipeline and applying slit-loss corrections. Our line luminosities are summarized in Table 3.

\subsection{Intrinsic UV Luminosity}
To understand the general property of our high-redshift sample, we calculated a magnification-corrected UV luminosity for each galaxy,
\begin{equation}
\frac{L}{L^*} = 10^{0.4(M^*_{\textrm{UV}}-M_{\textrm{UV}})},
\end{equation}
where M$^*_{UV}$ is the redshift-dependent UV absolute magnitude of a typical galaxy at a given redshift interval taken from \citet{bouwens2015uv}, and 
\begin{multline}
M_{\textrm{UV}} \approx M_{\textrm{FUV}} = m_{\textrm{F160W}}+\text{2.5log$_{10}$($\mu$)} -\\5(\text{log$_{10}$}(10^6d_L)-1)+\text{2.5log$_{10}$(1+$z$)}+0.12,
\end{multline} where $m_{\textrm{F160W}}$ is the apparent magnitude in F160W, $d_L$ is the luminosity distance, and $\mu$ is the magnification. First we perform a $k$-correction to move from the observed-frame F160W band to the rest-frame GALEX FUV band. Then we move from GALEX NUV to GALEX FUV by adding 0.12 to $M_{\textrm{FUV}}$ (the average $M_{\textrm{FUV}}$-$M_{\textrm{NUV}}$ color of $\sim$500 galaxies at $z>3.5$). We assume the rest-framed F160W response curve at the median redshift ($\sim 6$) of our sample is similar to that of the observed-frame GALEX NUV. Our intrinsic luminosities for objects with photometric or spectroscopic redshifts above $z=4$ are presented in Figure 4. They range from $3.5L^*$ down to  0.001$L^*$ with a median value of $0.03L^*$. The median value among all of our detections is $0.13L^*$.

\begin{figure}
  \centering
  \includegraphics[width=\linewidth]{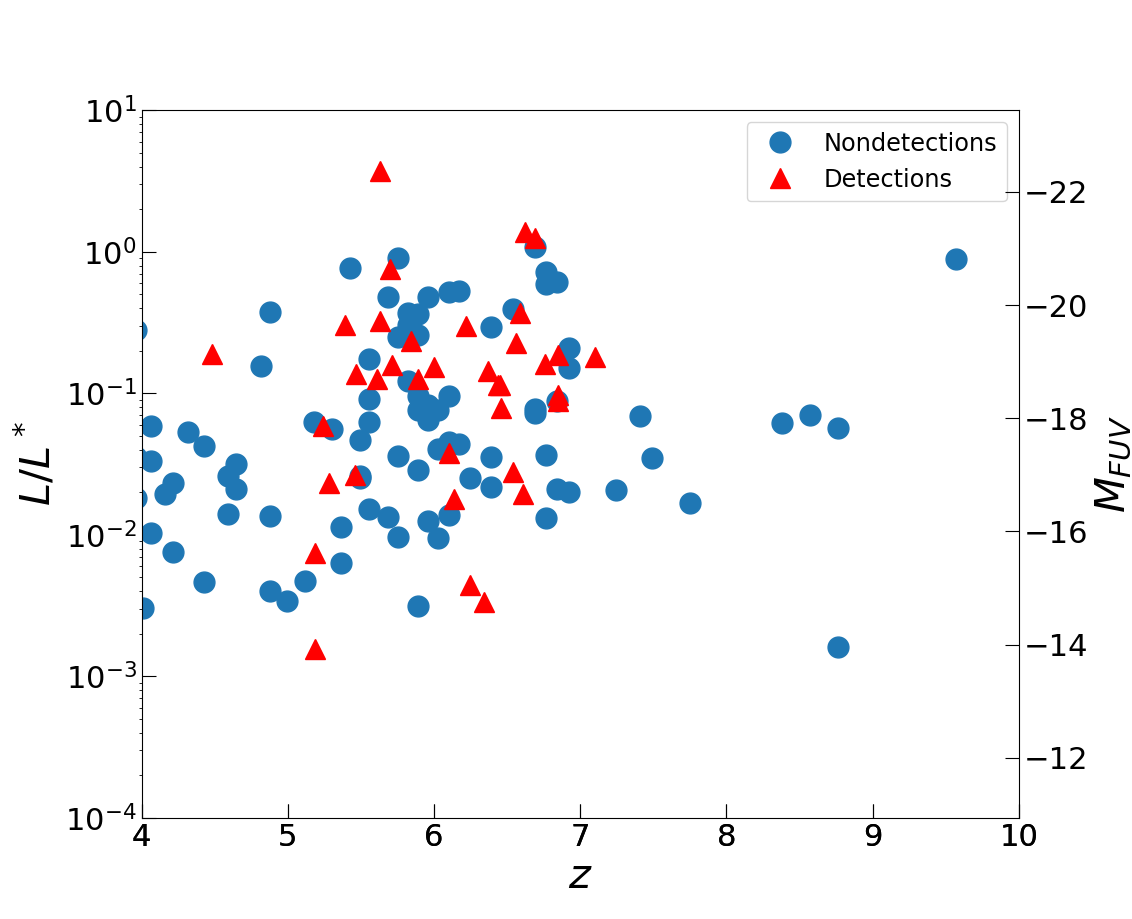}
  \caption[width=\linewidth]{Magnification corrected $L/L^*$ and $M_{\textrm{FUV}}$ values for objects in our sample with $z_{\textrm{phot}}$ or $z_{\textrm{spec}}$ between $z=4$ and $z=10$. $M^*_{\textrm{UV}}$ for the various redshifts were taken from \citet{bouwens2015uv}. The median $L/L^*$ is 0.03 for our full sample, 0.07 for only those objects with $z_{\textrm{phot}}$ or $z_{\textrm{spec}} > 4$, and 0.13 for our detections.}
\end{figure}

The rest-frame equivalent width (EW = $f_{\textrm{Ly}\alpha}/(f_{\lambda}$(1+$z$)) for each night of observation was calculated by dividing the weighted line flux found above by the flux density, 

\begin{equation}
f_{\lambda\textrm{F105W}} = 10^{-0.4(m_{\textrm{AB,F105W}}+48.6)}c/\lambda^2,
\end{equation}
where $\lambda = 10438.9$Å is the effective wavelength of the F105W (or adjacent redward) filter. The F105W band is redward of the emission line in most cases, however, in some cases the F125W magnitude was used. If neither of these magnitudes were measured, the F140W magnitude was used. For these cases, we verified the magnitude of our LAEs did not change significantly between the filters. The EW and errors are summarized in Table 3. Figure 5 presents a histogram of our rest-frame equivalent widths for all of our detections.

\begin{figure}
  \includegraphics[width=\linewidth]{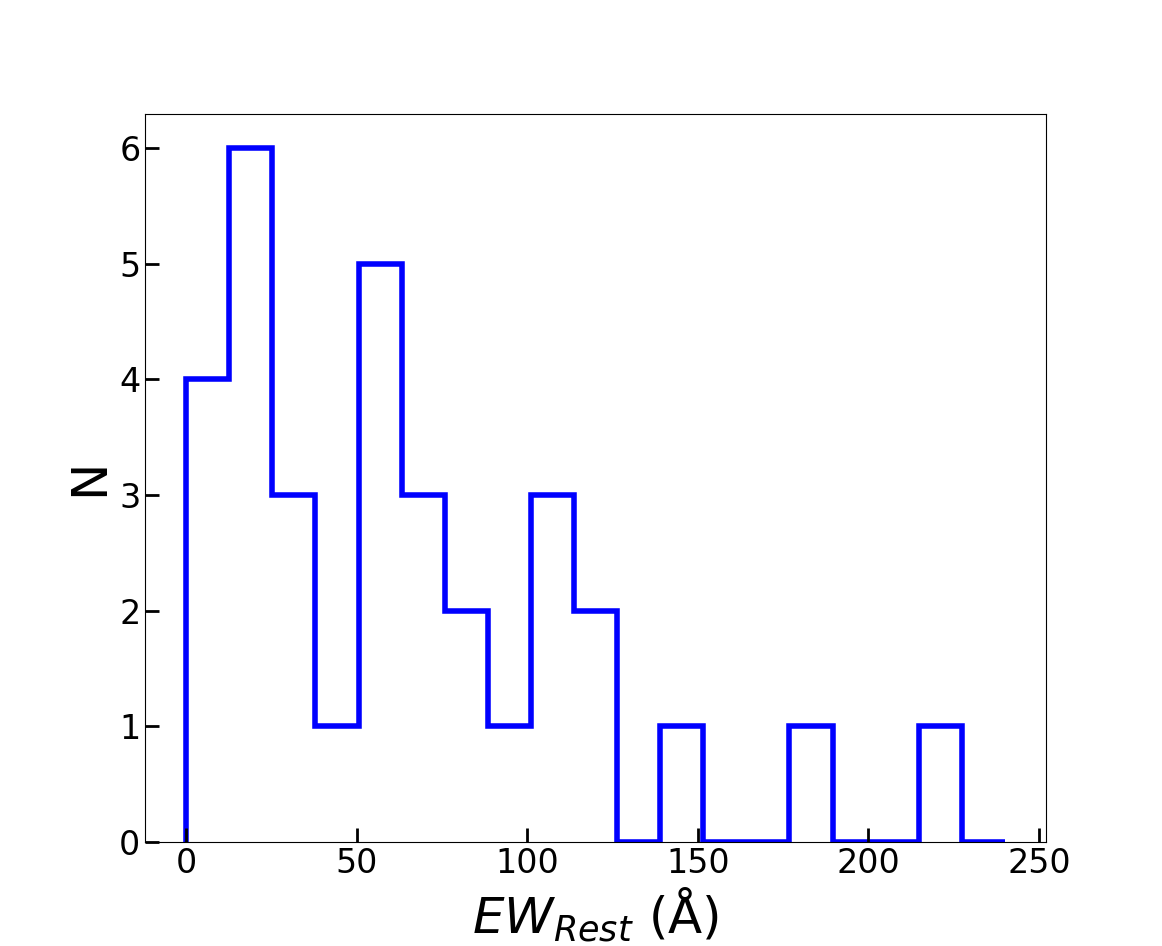}
  \caption{Histogram of the rest-frame EWs for our detections. These are calculated by averaging the 1/noise-weighted EWs, where the noise is the standard deviation in the flux from the inverse variance flux spectrum for a night of observation. EWs are summarized in Table 3.}
\end{figure}

We used the $1\sigma$ flux error for each night of observation to calculate the EW limit. Our median $1\sigma$ flux error is $1.6\times10^{-18}$ erg/s/cm$^2$ and our mean and median rest-frame $1\sigma$ EW limits for all detections are 17Å and 13Å respectively.

\begin{figure}[H]
  \includegraphics[width=\linewidth]{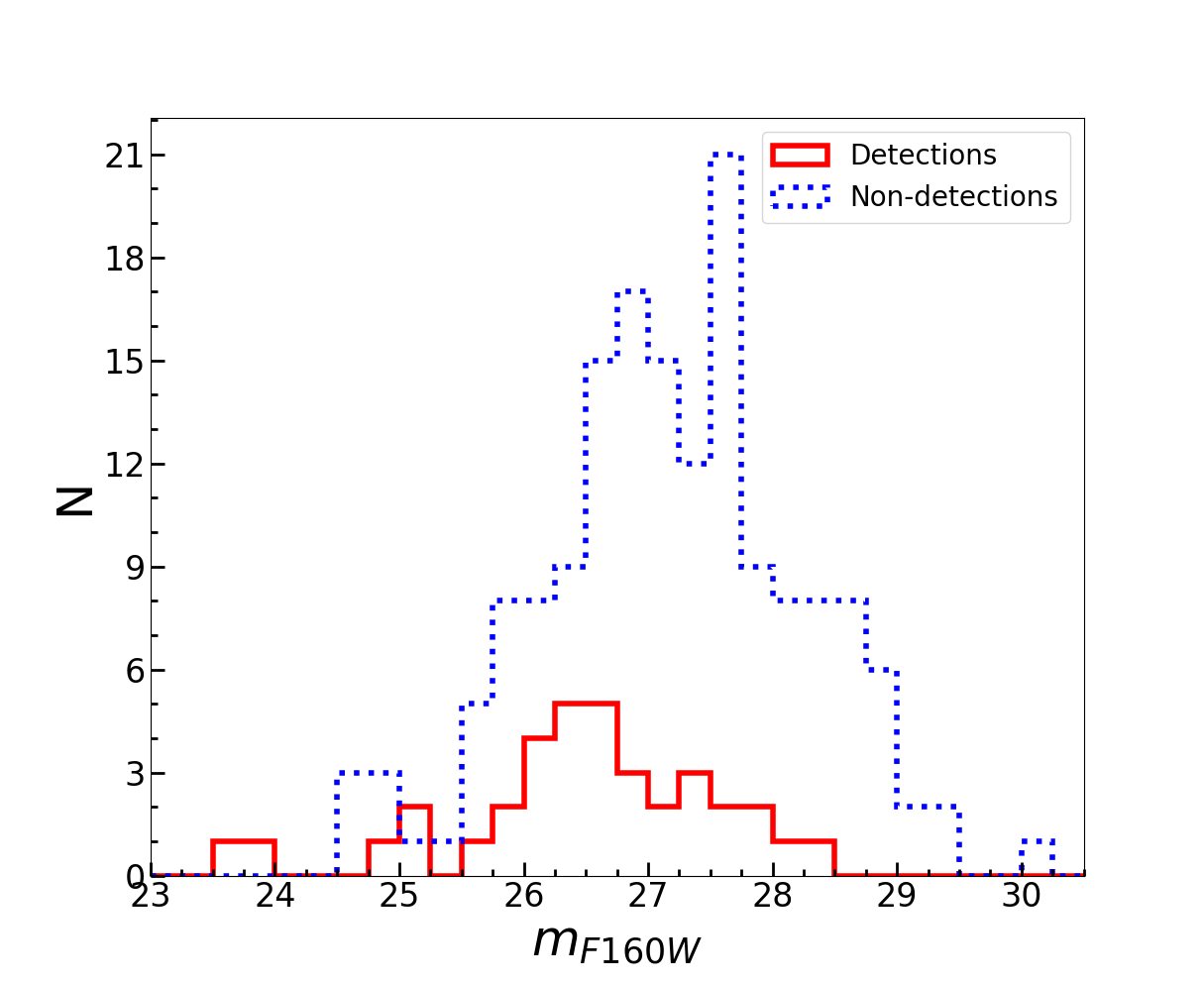}
  \caption{Histogram of F160W apparent magnitudes ($m_{\textrm{F160W}}$) for our sample of high-redshift objects, separated by detections and non-detections.}
\end{figure}

\begin{figure}[H]
  \includegraphics[width=\linewidth]{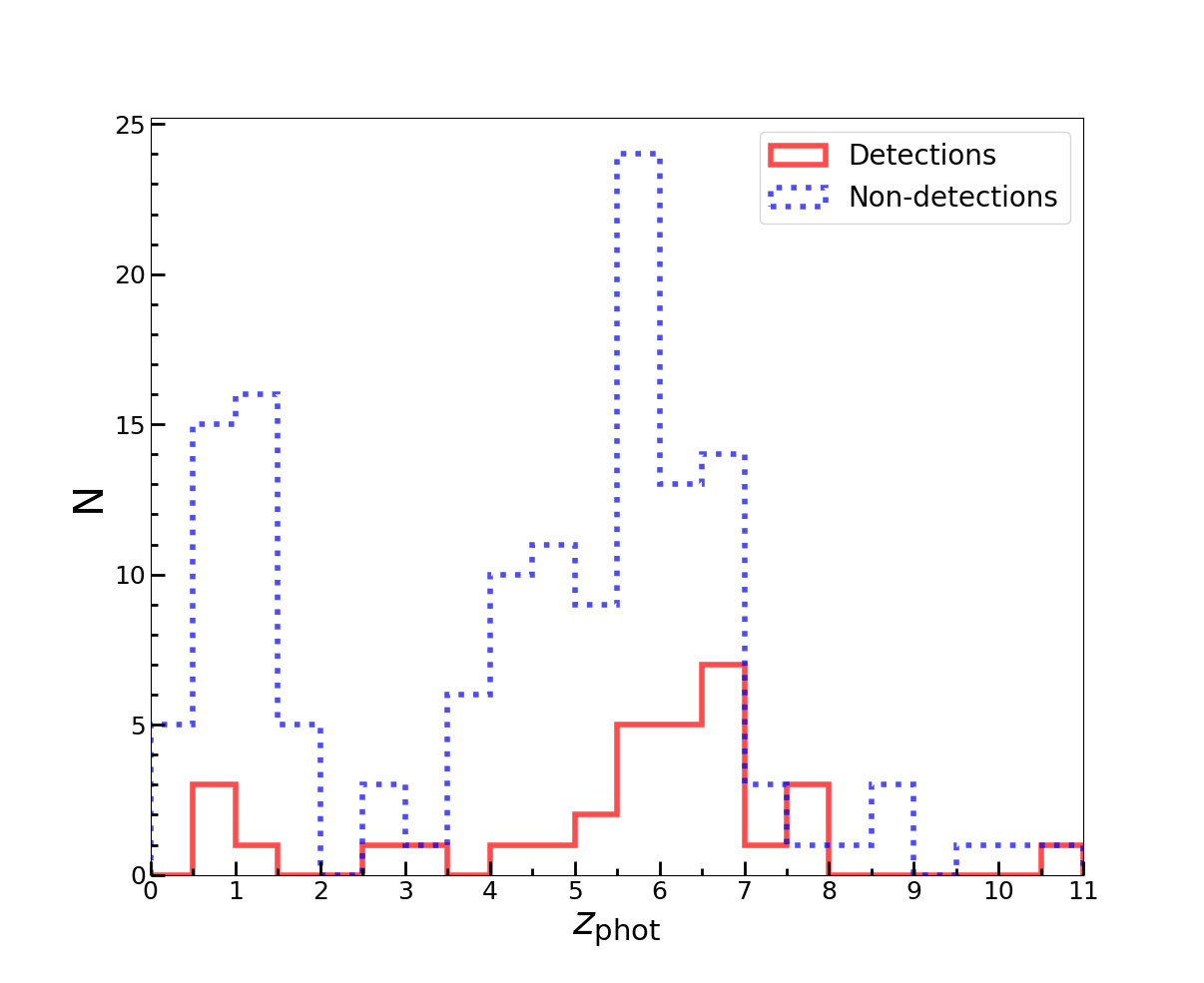}
  \caption{Histogram of $z_{\textrm{phot}}$ values without priors corresponding to the dominant peaks from the full P($z$)s. These are separated by detections and non-detections.}
\end{figure}

\section{Results}
After our visual and automated search methods, we found 19 objects with significant Ly$\alpha$ emission and an additional 17 objects with likely Ly$\alpha$ emission. In Table 3 we summarize the properties of our detections. Q4 detections are listed first, followed by Q1-Q3 detections. A number of these clusters were observed with MUSE for lens modeling and strong lensing analyses \citep{mahler2017strong,caminha2019strong,caminha2017refined,lagattuta2019probing,lagattuta2017lens}, however, we only found one counterpart among our DEIMOS observations. Additionally, some of out detections were spectroscopically observed by other groups. Details can be found in the table footnotes.

In Figure 6 we plot the apparent magnitudes in F160W now separated by detection and non-detection. We were able to detect Ly$\alpha$ in objects with magnitudes as faint as $m_{\textrm{F160W}}\sim28.5$. In Figure 7 we plot $z_{\textrm{phot}}$ values again separated by detections and non-detections. Two Q4 detections, IDs 2129.22 and 2214.1, have $z_{\textrm{phot}} \lesssim 5$. Object 2129.22 has a secondary peak at $z>5$ and has previously been confirmed spectroscopically as a multiple image \cite{huang2016detection}. For object 2214.1, although the photometric redshift strongly suggests $z_{\textrm{phot}}<z_{\textrm{spec}}$, the shape of the line is convincingly Ly$\alpha$ as it has a long tail on the red end.

In Figure 8 we present a plot of the photometric redshift against the spectroscopic redshift determined from the wavelength of the Ly$\alpha$ detection. 

\begin{figure}[H]
  \includegraphics[width=\linewidth]{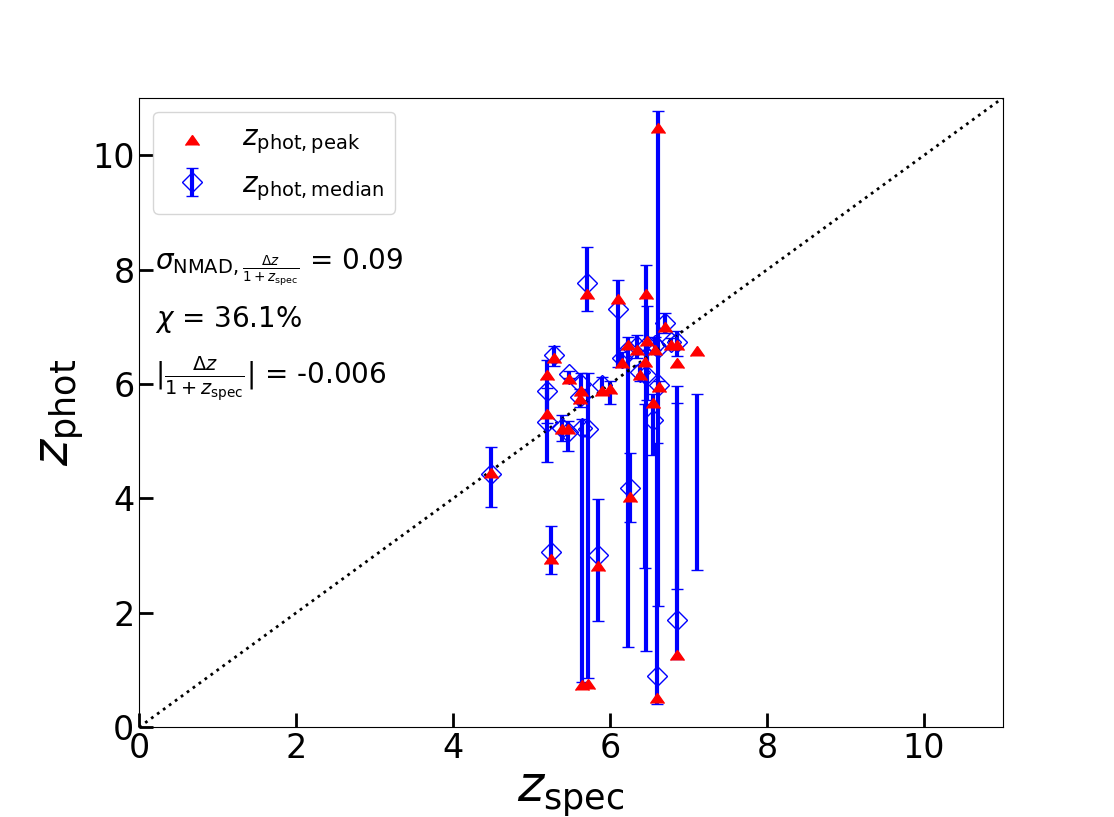}
  \caption{Photometric redshifts versus spectroscopic redshifts. The blue circles are the median P($z$) values and the error bars correspond to the lower and upper 1$\sigma$ values from the P($z$). The median 1$\sigma$ values were estimated from the 2$\sigma$ values for the four CLASH detections. In red are peak values of the P($z$)s. Also shown in black is the line where $z_{\textrm{phot}} = z_{\textrm{spec}}$. We found an NMAD scatter of 0.09, a catastrophic outlier percent of 36.1\%, and a bias of -0.006. The catastrophic outlier is calculated as the fraction of objects where $z_\textrm{spec}$ is $>$15\% away from $z_{\textrm{phot,peak}}$.}
\end{figure}

\subsection{LAE Fraction}
In this section we investigate how the fraction of LAEs changes with redshift in our sample. This analysis is particularly useful when looking at two intrinsically similar samples with a small difference in redshift such that we can approximate to the accuracy of this test that the physics of the ISM does not evolve appreciably. The results from the LAE fraction test can give us a hint about how the neutral hydrogen fraction is changing in the IGM and CGM. A full Bayesian analysis involving radiative transfer is needed for a more complete understanding (e.g., as in \citealt{mason2018universe,hoag2019constraining,mason2019inferences}). This modeling will be done in the future for our sample using the rest-frame Ly$\alpha$ equivalent width measurements for our detections, and the equivalent width limit spectra for our non-detections (Lexmaux et al., in prep.). 

In order to calculate the LAE fraction with reasonable accuracy, we must statistically incorporate the photometric information of the non-detections. A simple calculation of the fraction of LAEs among LBGs is not sufficient, as it does not include uncertainty on the redshifts of the non-detections, nor uncertainty on the EWs for our detections, nor take into account our observational EW and magnitude limits. To account for these uncertainties, we ran 1000 Monte Carlo (MC) trials on the reconstructed P($z$)s for our non-detections for two different redshift bins, $5\leq z\leq6$ and $6<z\leq7$. For each trial, we sampled one redshift from each of the full P($z$)s of the non-detections and counted the number of galaxies that returned a trial redshift that corresponded to a Ly$\alpha$ observed-frame wavelength that fell in either the lower or upper redshift bin. This sample comprised our non-detections. For each trial, we then Gaussian sampled the EWs of our detections with their associated uncertainties and counted how many detections returned EW$>$25Å in each redshift bin. We calculated the LAE fraction as $n_{\textrm{LAE}}$/$n_{\textrm{Total}}$. To produce a more comparable estimate of the LAE fraction, we adjusted our full high-redshift candidate sample with a series of cuts motivated by our observations and observations from the literature.

First, we made an EW cut at 25Å for the LAE sample, as this is the canonical limit from the literature for Ly$\alpha$ emission (e.g. \citealt{pentericci2018candelsz7,mason2018universe}). We then made a magnitude cut at $m_{\textrm{AB}}=26.8$ for the high-redshift sample. For our average observing conditions and exposure times, this is the magnitude at which we would expect to detect Ly$\alpha$ with EW$\geq$25Å. Our median flux density error is $1.6\times10^{-18}$erg/s/cm$^2$, which corresponds to an EW limit of $\sim$8Å for a galaxy of $m_{\textrm{AB}}=26.8$. If we detect Ly$\alpha$ from a galaxy at our EW cut of 25Å, this would correspond to a 3$\sigma$ detection, which matches the S/N=3 limit we used in our search. These cuts left us with 13 detections and 40 non-detections. All of the Q1 detections except for one were removed from the sample and more than 3/4 of our Q2 and Q3 detections were also removed. The remaining sample of detections comprised mainly of Q4 detections.

It is important to note that three of our detections were previously reported spectroscopically by various groups. These are IDs 2129.31 \citep{schmidt2016grism}, 370.14 \citep{hu2002redshift}, and 0717.25 \citep{vanzella2014characterizing,treu2015grism} in Table 3. Additionally, a number of our detections and non-detections were observed by GLASS, however only two detections were confirmed at wavelengths consistent with those in GLASS. Objects with GLASS detections were given a higher priority when designing the mask. The effects of removing these objects from the sample are discussed in Section 4.2.

Our full sample of detections consists of two multiply-imaged systems. One of the systems is triply-imaged \citep{huang2016detection} and the other is doubly-imaged. See Appendix C for a brief discussion of our two multiply-imaged systems. In the analysis that follows, multiple images of a galaxy were combined and treated as a single detection. We do not expect, on average, to be targeting more multiply-lensed systems among our detections than among our non-detections. To correct for possible multiple images among our non-detections, we randomly removed a fraction of the non-detections from each MC trial corresponding to the fraction of multiple images among our detections. The fraction was chosen randomly between 0.77 (1/13) and 0.14 (2/14) because one detection from the triply-imaged system did not make the magnitude cut, and the another detection from the doubly-imaged system did not make the EW cut, but could make the cut when incorporating the error. For this reason, we picked a random number from a uniform distribution between 0.077 and 0.14 when determining the fraction of multiple images among our non-detections.

We ran the MC analysis on our final sample and found a LAE fraction of 0.26$\pm0.04$ ($z\sim5.5$) and 0.30$\pm0.04$ ($z\sim6.5$), where the adopted LAE fraction is the median of the distribution of LAE fractions for our Monte Carlo trials. The error bars are a product of our statistical treatment of the LAE fraction, taking into account redshift uncertainties for the non-detections and EW uncertainties for the detections, with the lower and upper error bars taken from the 16$^\textrm{th}$ and 84$^\textrm{th}$ percentiles of the LAE fraction for our 1000 Monte Carlo trials.

We also split our sample by the median intrinsic UV luminosity of our detections, $0.13L^*$, to see if the incidence of galaxies emitting Ly$\alpha$ depended on intrinsic UV luminosity. We performed the same MC analysis on the two redshift bins for the fainter and brighter samples. For our fainter sample we found a LAE fraction of 0.29$^{+0.08}_{-0.04}$ ($z\sim5.5$) and 0.43$^{+0.14}_{-0.10}$ ($z\sim6.5$), and for our brighter sample, a LAE fraction of 0.22$^{+0.06}_{-0.02}$ ($z\sim5.5$) and 0.20$^{+0.02}_{-0.03}$ ($z\sim6.5$). Our results for the full, brighter, and fainter samples are plotted in Figure 9, along with results from the literature. It is important to note though, that the results from the literature are from samples with brighter intrinsic luminosities (except for \citealt{hoag2019constraining} and \citealt{mason2019inferences}), so a direct comparison cannot be made. In addition, the fact that the magnitude distribution of the non-detections and detections are not the same, and other selectional operational definition effects make comparison difficult.

\begin{figure*}[!htbp]
  \includegraphics[width=\linewidth]{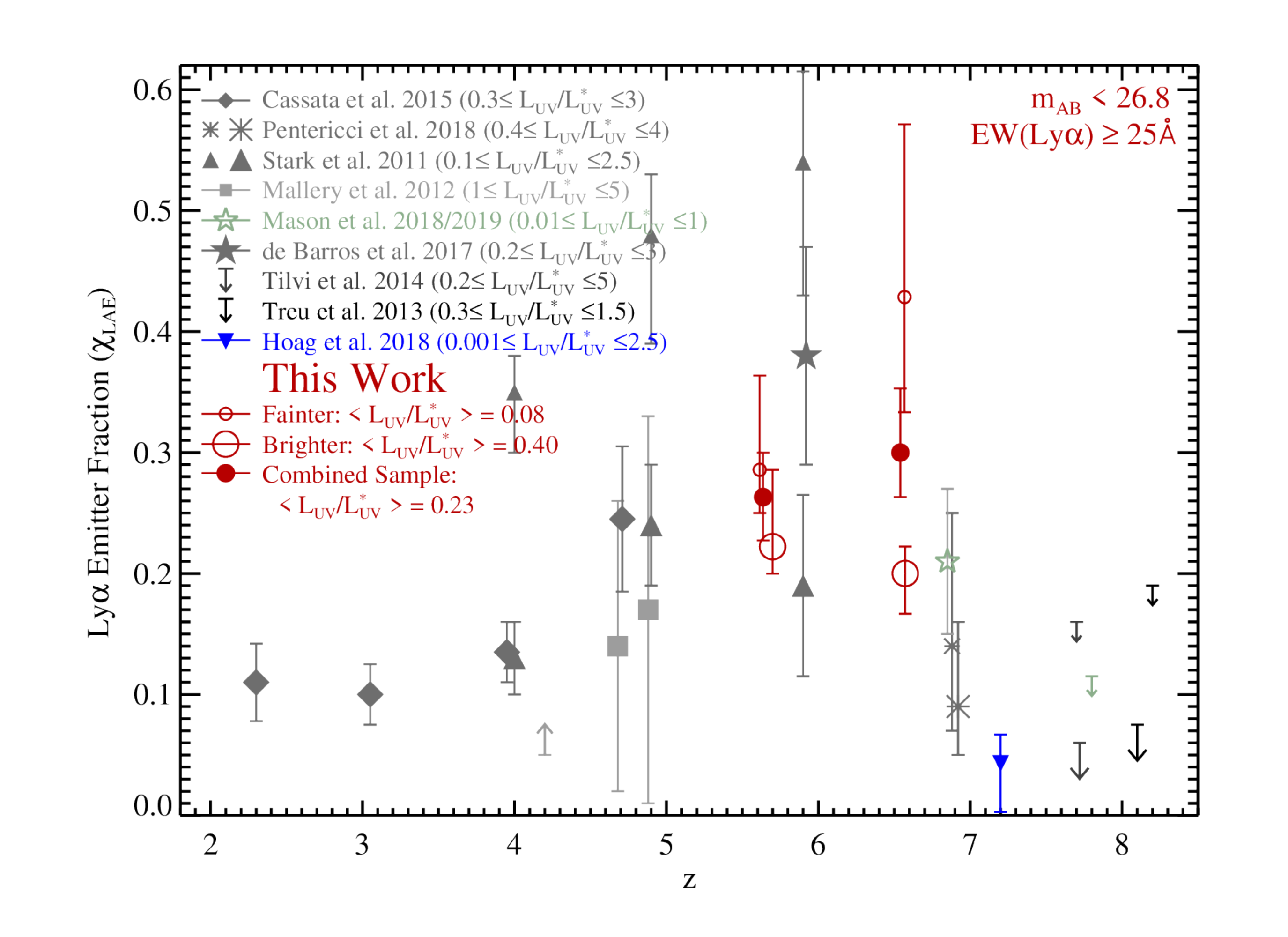}
  \caption{LAE fraction results after making a magnitude cut at $m_{\textrm{AB}}=26.8$ and an EW cut at $25$Å. The sample is also binned by intrinsic UV luminosity. Filled circles are the entire sample, small open circles are for $L/L^*<0.20$, and large open circles are for $L/L^*>0.20$, where 0.20 is the mean $L/L^*$ for our combined sample after the cuts. Median $L/L^*$ values are shown for our fainter, brighter, and combined sample. The fractions and error bars are the medians, and 16th and 84th percentiles of the trials respectively. Also shown are results from the literature.}
\end{figure*}

\subsection{Observational, Conceptual, \& Sample Selection Effects}
In this section we look into the selection effects to determine if the general result presented above can be recreated.

We first looked at what happened if we were to use all 36 detections, with equivalent widths ranging from $\sim$1Å to $\sim$400Å and apparent magnitudes from 24.1 to 28.5, and all 162 non-detections. This selection would be observationally motivated, as we do not expect observational effects to be different for high-redshift candidates at $z\sim5.5$ than for those at $z\sim6.5$. We calculated a LAE fraction of 0.31$^{+0.03}_{-0.02}$ ($z\sim5.5$) and 0.40$^{+0.04}_{-0.03}$ ($z\sim6.5$), which is still consistent with a LAE fraction that does not decrease at higher redshifts. These values are similar to those found above, differing by 1.1$\sigma$ and $2\sigma$ for the lower- and higher-redshift samples, respectively. We performed the same MC analysis with fainter and brighter magnitude cuts of 27.3 and 26.3 respectively, corresponding to the 1$\sigma$ spread of our observing depths (see Section 2.5). Again, we found the lower luminosity galaxies had an equal or greater LAE fraction than the brighter sample at a given redshift, and the LAE fraction did not drop significantly for either of the combined samples from $z\sim5.5$ to $z\sim6.5$.

In another analysis, we counted the multiply-imaged detections from our sample as individual detections and did not remove a fraction of our non-detections from the analysis, as we cannot quantify the number of multiply-imaged non-detections with certainty. We again found a slope of the LAE fraction with redshift that was statistically consistent with flat, similar to previous results, and fainter galaxies again having a slightly higher fraction than brighter galaxies.

Finally, because we assigned higher priorities to GLASS objects with detections, we decided to remove these objects and see how the LAE fraction changed from selection effects. Removing the GLASS objects did not have a significant effect on the fainter objects, however, we did see different drop in the LAE fraction for the brighter objects from $z\sim5.5$ to $z\sim6.5$. To see if this decrease was an effect of small number statistics, we ran a jackknife on our sample by first creating a sub-sample with the same number of objects as we have GLASS-selected objects. We removed this sub-sample from our original sample and ran the same Monte Carlo analysis as before. We performed this 1000 times and found a $\sim1.5\sigma$ difference between the jackknife LAE fraction results and the GLASS-excised LAE fraction results. Though we can't rule out a statistical rarity, we can still look into what is causing this decrease in LAE fraction for our bright GLASS-excised sample. This decrease is likely due to two properties of GLASS. First, the \textit{HST} grism is only sensitive down to 8000Å, whereas our DEIMOS observations were sensitive down to 7500Å or lower for the 600ZD and 830G gratings. More importantly, GLASS observed intrinsically brighter galaxies, with a $1\sigma$ EW limit of $\sim$120Å at $m_{AB} = 26.8$. This explains why the results for our higher redshift, brighter sample changed, but our entire sample did not change significantly.

\subsection{Interpretation}
The consistent results from our multiple analyses tell us that our results related to the redshift evolution of the LAE fraction for the full sample and sub-samples of differing brightnesses are robust. We consistently see a statistically flat or gradually rising slope in the LAE fraction from $z\sim5.5$ to $z\sim6.5$. Additionally, we find a different LAE fraction for our UV fainter and UV brighter samples. In our complete analysis presented in Section 4.1, there is a 0.9$\sigma$ difference between the LAE fraction for the brighter and fainter samples at $z\sim5.5$ and a 2.3$\sigma$ difference between the two fraction at $z\sim6.5$. Our fraction for the brighter sample agrees better with what is found in the literature. Our luminosity bin cutoff of $0.13L^*$ is at the lower extreme of all of the samples from the literature except for \citet{mason2018universe,mason2019inferences}, and \citet{hoag2019constraining}, which covers a similar luminosity range as we do. Comparing our results with just the \citet{hoag2019constraining} sample, we find the LAE fraction must drop precipitously between $z\sim6.5$ and $z\sim7.5$, a difference in cosmic time of only $\sim$100Myr. In this window of time, the environment surrounding low luminosity galaxies is changing rapidly, allowing more Ly$\alpha$ photons to escape. For a quantitative comparison of the brighter objects, \citet{pentericci2018candelsz7} found a LAE fraction at $z=6$ for galaxies between $-20.25 < M_{\textrm{UV}} < -18.75$ with rest-frame Ly$\alpha$ EWs$>$25Å of 0.14$^{+0.11}_{-0.07}$, which is in agreement with our value of 0.18$^{+0.04}_{-0.02}$ ($z\sim6.5$) for our brighter sample with similar properties. The difference between our fainter LAEs and brighter galaxies in the literature and in our own sample, indicates that lower luminosity galaxies are collectively emitting more Ly$\alpha$ emission than their brighter counterparts, particularly at $z\sim6.5$. This discrepancy is possibly due to differences in $f^{\textrm{Ly}\alpha}_{\textrm{esc}}$ between the two samples.

Assuming the intrinsic Ly$\alpha$ EW distributions are the same at $z=5.5$ as at $z=6.5$, our results support the Ly$\alpha$ luminosity function results from field surveys found at these redshifts, which show a suppression at the bright end at $z\sim6.5$ \citep{kashikawa2011completing,dressler2015confirmation,santos2016lyalpha}. Furthermore, the Ly$\alpha$ luminosity functions at these redshifts have steep faint-end slopes, meaning there were far more intrinsically faint galaxies emitting Ly$\alpha$ than intrinsically bright galaxies. If the LAE fraction for our fainter galaxies is indeed rising, then these galaxies potentially played an important role in reionization. That being said, there is some evidence from Ly$\alpha$ luminosity functions from objects with larger line luminosities than the objects presented here, that the number density of LAEs drops smoothly across all Ly$\alpha$ luminosities from $z\sim6$ to $z\sim7$ \citep{hu2019ly}. Further work is needed to make a conclusive statement about the the evolution of the number density of objects with faint Ly$\alpha$ luminosities.

\section{Conclusion}
Using Keck/DEIMOS, we observed 198 high-redshift candidates photometrically selected to be between $5\lesssim z \lesssim 7$. Of these we found 36 objects with Ly$\alpha$ emission. We ran Monte Carlo trials on the full P($z$)s of our detections and non-detections after making a cut at $m_{\textrm{AB}}<26.8$ and EW$>$25Å, to generate a LAE fraction for two redshift bins at $z\sim5.5$ and $z\sim6.5$. For our entire sample, we found a LAE fraction of 0.26$\pm0.04$ ($z\sim5.5$) and 0.30$\pm0.04$ ($z\sim6.5$). We also split our sample into a brighter and fainter sample based on our median intrinsic UV luminosity of our detections, $0.13L^*$. We calculated a LAE fraction of 0.29$^{+0.08}_{-0.04}$ ($z\sim5.5$) and 0.43$^{+0.14}_{-0.10}$ ($z\sim6.5$) for our fainter sample, and 0.22$^{+0.06}_{-0.02}$ ($z\sim5.5$) and 0.20$^{+0.02}_{-0.03}$ ($z\sim6.5$) for our brighter sample. In general, our full LAE fraction is consistent with findings from other groups out to $z=6$, however, at higher redshifts and lower luminosities, our results diverge from what other groups have found with galaxies with brighter luminosities as well as those in our sample.

When we compare our sample to the parallel MOSFIRE sample at $7\lesssim z\lesssim8.2$ \citep{hoag2019constraining}, which spans a similar luminosity range, we see a sharp drop in the LAE fraction from $z\sim6.5$ to $z\sim7.5$, indicating a rapid evolution of the environment surrounding these galaxies. The results from our brighter sample, on the other hand, indicate the LAE fraction does not drop as significantly at $z\sim6.5$, and remains relatively flat between $z\sim5.5$ and $z\sim6.5$. It is important to keep in mind, comparisons cannot be made with most of the samples from the literature as they have different intrinsic properties.

To confirm our results and to infer the neutral hydrogen fraction, we will input our detections and non-detections into a Bayesian framework (\citealt{mason2018universe,mason2019inferences,hoag2019constraining}, Lemaux et al. in prep.). A base sample is crucial for the correct interpretation of $z\sim7$ and above measurements. We can then make a comparison between the neutral hydrogen fraction for our sample between $5\lesssim z\lesssim7$ and the MOSFIRE sample at $z\sim7.5$ and study the tail-end of reionization.

The data presented herein were obtained at the W. M. Keck Observatory, which is operated as a scientific partnership among the California Institute of Technology, the University of California, and the National Aeronautics and Space Administration. The Observatory was made possible by the generous financial support of the W. M. Keck Foundation. We wish to recognize and acknowledge the very significant cultural role and reverence that the summit of Maunakea has always had within the indigenous Hawaiian community.  We are most fortunate to have the opportunity to conduct observations from this mountain. Observations were also made with the \textit{Hubble Space Telescope}, operated by the Space Telescope Science Institute, and the \textit{Spitzer Space Telescope}, operated by the NASA Jet Propulsion Laboratory. Data from SDSS were also used in a cursory fashion. Funding for SDSS-III has been provided by the Alfred P. Sloan Foundation, the Participating Institutions, the National Science Foundation, and the U.S. Department of Energy Office of Science. The SDSS-III web site is http://www.sdss3.org/. SDSS-III is managed by the Astrophysical Research Consortium for the Participating Institutions of the SDSS-III Collaboration.

This research is supported in part by NSF through grant NSF AST-1815458 to M.B. and grant AST-1810822 to T.T., and by NASA through grant NNX14AN73H to M.B. and grant HST-GO-13459 (GLASS) to T.T.. C.M. acknowledges support provided by NASA through the NASA Hubble Fellowship grant HST-HF2-51413.001-A awarded by the Space Telescope Science Institute, which is operated by the Association of Universities for Research in Astronomy, Inc., for NASA, under contract NAS5-26555.

We would like to acknowledge the ASTRODEEP and CLASH teams for their work on generating photometric catalogs for many of these clusters.

Finally we would like to thank Debora Pelliccia for her insightful comments in our discussions about LAE fractions, Patty Bolan, and Pratik Gandhi.

\bibliography{refs}
\bibliographystyle{apj}

\section*{Appendix}
\section*{A. ASTRODEEP and CLASH Photometric Catalogs}
Four of our Ly$\alpha$ candidates and a number of our bright objects for flux calibration did not exist in the ASTRODEEP photometric catalogs. For these cases, we turned to the CLASH photometric catalogs. In this section, we present a brief comparison of the photometry in these two catalogs, both in isophotal magnitudes. For objects of all brightnesses in F814W, the ASTRODEEP catalog in general had slightly dimmer magnitudes, particularly for the brighter offsets. Table 5 summarizes the differences between the catalogs. The offsets are small and are not a concern for our photometry.

\subsection*{B. Ly$\alpha$ Detections}

\clearpage
\begin{turnpage}
\def\arraystretch{2}
\begin{table}[H]
\tiny
\vspace{-1cm}
\hspace{-1cm}
\caption{Detections.}
\begin{tabular}{| l l l l l l l l l l l l l l l l|} \hline

Cluster.ID & Mask & RA & Dec & $\lambda_{Ly\alpha}$(Å) & S/N &EW (Å)&\parbox{1.5cm}{\centering L$_{\textrm{Ly}\alpha}$ \\ ($\times10^{42}$erg/s)}& Asym.&Ly$\alpha$ Quality& $z_{\textrm{spec}}$& P($z$)$_{\textrm{Peak}}$& P($z$) & $m_{\textrm{F160W}}$& $M_{\textrm{FUV}}$ & $\mu$\\ \hline \hline
1347.36	&\parbox{1cm}{\centering miki13DG\\miki13DG2\\134717A1}&206.903015	&$-$11.750369    &9083.0 $\pm$ 1.1	&16.0&112$\pm$21&2.4$\pm$0.2&1.2&Q4&$6.471^{+0.003}_{-0.001}$&6.844&6.690$^{+0.68}_{-0.97}$&27.4 $\pm$ 0.3&-18.1&4.1$^{+0.2}_{-0.1}$ \\  \hline
1347.47	&\parbox{1cm}{\centering miki13D\\miki13DG\\miki13DG2}&206.900859	&$-$11.754209    &9447.1 $\pm$ 0.8$^{*,3,4,6}$	&5.4 &55.4$\pm$10.2&2.7$\pm$0.4&1.1$^*$&Q4&$6.771\pm0.001$&6.766&6.741$^{+0.07}_{-0.09}$&26.1 $\pm$ 0.1&-18.9&5.2$^{+0.3}_{-0.2}$\\ \hline
2129.22    &\parbox{1cm}{\centering M2129\_D3\\M2129\_D4}&322.350939	&$-$7.693331	    &9538.7 $\pm$ 0.6$^2$	&15.1 &99.7$\pm$17.8&5.2$\pm$0.6&5.5&Q4&$6.846\pm0.001$&1.353&1.869$^{+4.1}_{-0.68}$&26.3 $\pm$ 0.1&-19.1&3.79$^{+0.02}_{-0.02}$\\ \hline
2129.28    &\parbox{1cm}{\centering 212917A1}&322.336347	&$-$7.696575	    &8057.7 $\pm$ 0.5	&6.2 &36.7$\pm$5.1&1.5$\pm$0.2&2.3&Q4&$5.628\pm0.001$&5.961&5.992$^{+0.20}_{-0.18}$&26.2 $\pm$ 0.2&-19.7&1.88$^{+0.02}_{-0.03}$\\ \hline
2129.31    &\parbox{1cm}{\centering M2129\_D3\\M2129\_D4\\212915A1\\212915B1}&322.353238	&$-$7.697444	    &9538.5 $\pm$ 0.8$^{*,2,6,7}$	&13.0 &57.9$\pm$11.4&2.5$\pm$0.4&3.6$^*$&Q4&$6.846\pm0.001$&6.766&6.726$^{+0.19}_{-0.24}$&26.8 $\pm$ 0.2&-18.2&5.40$^{+0.05}_{-0.04}$\\ \hline
2129.36    &\parbox{1cm}{\centering M2129\_D3\\M2129\_D4\\212915A1\\212915B1}&322.353941	&$-$7.681644	    &9538.8 $\pm$ 0.5$^{*,2,6}$	&6.5 &82.3$\pm$36.7&1.9$\pm$0.5&3.5$^*$&Q4&$6.846^{+0.001}_{-0.000}$&6.456&[0.776,7.309]&28.3 $\pm$ 0.3&-18.3&1.210$\pm0.001$\\ \hline
1423.17    &\parbox{1cm}{\centering C14215A1\\C14215A2}&215.972591	&24.072661	    &9057.2 $\pm$ 0.6$^{*,6}$	&12.0 &125$\pm$23&1.8$\pm$0.3&0.3$^*$&Q4&$6.450\pm0.001$&6.478&[1.339,7.087]&27.8	$\pm$ 0.3&-18.6&1.41$^{+0.04}_{-0.04}$\\ \hline
1423.26    &\parbox{1cm}{\centering miki14D\\miki14D2\\C14215A1\\C14215A2\\142317A1\\142317A2}&215.935869	&24.078415	    &8785.0 $\pm$ 0.2$^{*,3,6,7}$	&15.7 &59.3$\pm$9.5&2.2$\pm$0.3&1.1$^*$&Q4&$6.226^{+0.001}_{-0.000}$&6.766&6.589$^{+0.23}_{-5.19}$&26.4	$\pm$ 0.1&-19.6&2.06$^{+0.08}_{-0.06}$\\ \hline
1423.38    &\parbox{1cm}{\centering miki14D\\miki14D2\\142317A2}&215.93638	&24.057093	    &7766.5 $\pm$ 1.4$^*$	&25.7 &42.2$\pm$4.6&1.8$\pm$0.1&1.2$^*$&Q4&$5.389^{+0.001}_{-0.002}$&5.302&5.237$^{+0.22}_{-0.23}$&25.9 $\pm$ 0.1&-19.9&1.93$^{+0.06}_{-0.06}$\\ \hline
370.14    &\parbox{1cm}{\centering A370\_D3n\\A370\_D4n}&39.978069	&$-$1.558956	&9205.7 $\pm$ 1.3$^1$   &23.2 &79.5$\pm$6.5&0.1$\pm$0.6&2.6&Q4&$6.572\pm0.001$&6.689&6.685$^{+0.14}_{-0.09}$&25.12	$\pm$ 0.04&-19.3&8.9$^{+1.1}_{-0.8}$\\ \hline
370.55    &\parbox{1cm}{\centering A37017B1\\A37017B2}&39.990193	&$-$1.5712087	&8040.5 $\pm$ 0.5   &11.7 &101$\pm$14&2.5$\pm$0.3&2.5&Q4&$5.614\pm0.001$&5.824&5.779$^{+0.13}_{-0.18}$&26.57 $\pm$ 0.09&-18.7&3.23$^{+0.07}_{-0.07}$\\ \hline
0416.17    &\parbox{1cm}{\centering 041615B1}&64.03651	&$-$24.0923	    &8501.8 $\pm$ 0.4	&20.0 &66.0$\pm$10.4&2.5$\pm$0.3&0.9&Q4&$5.993^{+0.001}_{-0.000}$&6.007&[5.453,6.255]&26.5 $\pm$	0.1&-19.1&2.59$^{+0.06}_{-0.08}$\\
\hline
0416.56    &\parbox{1cm}{\centering 041615B1}&64.047848	&$-$24.062069	&8687.4 $\pm$ 1.3	&10.7 &105$\pm$23&1.6$\pm$0.3&6.5&Q4&$6.146\pm0.002$&6.463&6.450$^{+0.12}_{-0.10}$&26.69	$\pm$ 0.07&-16.6&22.6$^{+2.4}_{-1.3}$\\ \hline
0717.15  &\parbox{1cm}{\centering 071715B1}&109.3923348	&37.7380827	    &7866.0 $\pm$ 0.8$^*$	&15.8 &16.9$\pm$2.0&3.9$\pm$0.3&0.8$^*$&Q4&$5.470^{+0.002}_{-0.001}$&6.172&6.167$^{+0.05}_{-0.06}$&23.77	$\pm$ 0.02&-18.8&37.9$^{+1.6}_{-2.1}$\\ \hline
0717.17    &\parbox{1cm}{\centering 071715B1}&109.3914431	&37.7670479	    &7851.0 $\pm$ 0.6$^*$	&5.8 &69.3$\pm$17.5&1.0$\pm$0.2&1.0$^*$&Q4&$5.458\pm0.001$&5.302&5.161$^{+0.19}_{-0.33}$&27.7 $\pm$	0.6&-17.2&4.0$^{+0.3}_{-0.3}$\\ \hline
0717.25    &\parbox{1cm}{\centering 071715B1}&109.4077276	&37.7427406	    &8964.7 $\pm$ 1.1$^{*,5,6}$	&20.0 &145$\pm$17&6.6$\pm$5.7&3.8&Q4&$6.374\pm0.002$&6.244&6.212$^{+0.12}_{-0.15}$&26.82	$\pm$ 0.06&-18.8&2.7$^{+0.6}_{-0.5}$\\ \hline
0717.53    &\parbox{1cm}{\centering 071715B1}&109.4128542	&37.7338042	    &8932.9 $\pm$ 1.1$^s$	&8.3 &17.0$\pm$4.9&1.6$\pm$0.3&0.8&Q4&$6.348^{+0.001}_{-0.002}$&6.689&6.673$^{+0.19}_{-0.22}$&26.68 $\pm$ 0.08&-14.7&34.8$^{+32.8}_{-8.7}$\\ \hline
0744.60    &\parbox{1cm}{\centering M744D\_D1\\M744D\_D2}&116.221028	&39.44103	    &8387.3 $\pm$ 0.5$^*$	&13.3 &56.6$\pm$10.8&1.5$\pm$0.3&0.7$^*$&Q4&$5.899^{+0.001}_{-0.000}$&5.961&5.989$^{+0.14}_{-0.14}$&26.4	$\pm$ 0.2&-18.7&3.8$^{+0.4}_{-0.2}$\\ \hline
2214.1    &\parbox{1cm}{\centering 221417B1\\221417B2}&333.7234552   &$-$14.0142965   &8324.3 $\pm$ 0.4   &23.5 &407$\pm$42&4.1$\pm$0.3&6.3&Q4&$5.847^{+0.001}_{-0.000}$&2.909&3.003$^{+0.99}_{-1.14}$&26.9	$\pm$ 0.2&-19.3&1.45$^{+0.02}_{-0.01}$\\ \hline
\end{tabular}
IDs 2129.36, 1423.17, 0416.17 correspond to detections with continuum flux densities measured from the CLASH catalog.
$^s$ \textrm{These are secondary objects.}\\
$^*$ \textrm{Indicates these emission lines are next to or on skylines. The centroid location errors and asymmetry values should not be trusted. N.B. The object in A2744 has an emission line covering many skylines and we were unable to get a formal uncertainty and asymmetry value.}\\
$^1$\textrm{Spectroscopically confirmed by and consistent with \citet{hu2002redshift}.}\\
$^2$\textrm{Spectroscopically confirmed by and consistent with  \citet{huang2016detection}.}\\
$^3$\textrm{Photometrically selected by \citet{huang2016spitzer}.}\\
$^4$\textrm{Spectroscopically confirmed by and consistent with \citet{bradavc2017alma}.}\\
$^5$\textrm{Spectroscopically confirmed by and consistent with \citet{vanzella2014characterizing}.}\\
$^6$\textrm{Spectroscopically observed by \citet{treu2015grism}.}\\
$^7$\textrm{Spectroscopically observed by \citet{hoag2019constraining}.}
\end{table}
\clearpage

\def\arraystretch{2}
\begin{table}[H]
\ContinuedFloat
\tiny
\vspace{-1cm}
\hspace{-1cm}
\caption{}
\begin{tabular}{| l l l l l l l l l l l l l l l l|} \hline
Cluster.ID & Mask & RA & Dec & $\lambda_{Ly\alpha}$(Å) & S/N &EW (Å)&\parbox{1.5cm}{\centering L$_{\textrm{Ly}\alpha}$ \\ ($\times10^{42}$erg/s)}& Asym.&Ly$\alpha$ Quality& $z_{\textrm{spec}}$& P($z$)$_{\textrm{Peak}}$& P($z$) & $m_{\textrm{F160W}}$& $M_{\textrm{FUV}}$ & $\mu$\\ \hline \hline
1347.28	&\parbox{1cm}{\centering 134717A1}&206.8750764	&$-$11.7588426	&9171.8 $\pm$ 0.8 &5.2 &58.4$\pm$12.7&1.1$\pm$0.2&1.7&Q3&$6.544^{+0.002}_{-0.001}$&5.756&5.377$^{+0.45}_{-0.63}$&26.9 $\pm$ 0.2&-17.1&15.6$^{+1.9}_{-1.9}$\\ \hline
1347.29	&\parbox{1cm}{\centering 134717A1}&206.8865703	&$-$11.7620709	&7529.8 $\pm$ 1.2$^*$	&5.6 &32.6$\pm$8.4&0.39$\pm$0.10&8.0$^*$&Q3&$5.194^{+0.001}_{-0.002}$&5.558&5.331$^{+0.60}_{-0.69}$&27.0 $\pm$ 0.3$^*$&-13.9&144$^{+38}_{-24}$\\ \hline
1347.45	&\parbox{1cm}{\centering 134717A1}&206.8816569	&$-$11.761483	&7530.2$^3$ $\pm$ 1.1$^*$	&6.1 &16.5$\pm$6.5&0.22$\pm$0.09&1.8&Q3&$5.195^{+0.002}_{-0.001}$&6.244&5.885$^{+0.53}_{-0.57}$&26.2 $\pm$ 0.2&-15.6&77.7$^{+6.9}_{-5.8}$\\ \hline
2129.5	&\parbox{1cm}{\centering 212915A1\\212915B1\\212917A1}&322.36419	&$-$7.701938	&9067.1 $\pm$ 1.4$^7$	&7.2 &116$\pm$30&1.7$\pm$0.4&1.3&Q3&$6.458\pm0.002$&7.664&6.161$^{+1.92}_{-4.83}$&27.4 $\pm$ 0.3&-18.3&2.7$^{+0.1}_{-0.1}$\\ \hline
1423.13	&\parbox{1cm}{\centering C14215A1\\C14215A2}&215.928816	&24.083906	&8144.3 $\pm$ 2.4$^{6,7}$	&5.7 &12.7$\pm$7.0&0.3$\pm$0.1&2.0&Q2&$5.699^{+0.004}_{-0.003}$&7.664&7.773$^{+0.62}_{-0.50}$&25.7 $\pm$ 0.2&-20.3&1.65$^{+0.04}_{-0.05}$\\ \hline
1347.4	&\parbox{1cm}{\centering 134717A1}&206.895685	&$-$11.754647	&9236.6 $\pm$ 1.6$^6$	&4.9 &76.1$\pm$23.4&1.2$\pm$0.4&2.3&Q2&$6.598^{+0.002}_{-0.003}$&0.596&0.893$^{+4.07}_{-0.49}$&27.2 $\pm$ 0.4&-19.8&1.42$^{+0.2}_{-0.1}$\\ \hline
1347.25	&\parbox{1cm}{\centering 134717A1}&206.8706306	&$-$11.753105	&8818.8 $\pm$ 1.5	&4.2 &25.1$\pm$6.9&0.5$\pm$0.1&3.5&Q1&$6.254^{+0.002}_{-0.001}$&4.113&4.189$^{+0.61}_{-0.61}$&26.3 $\pm$ 0.3&-15.0&177$^{+40}_{-38}$\\ \hline
1347.39	&\parbox{1cm}{\centering 134717A1}&206.8886406	&$-$11.7542822	&7642.2 $\pm$ 0.8	&3.2 &8.48$\pm$4.57&0.2$\pm$0.1&0.6&Q1&$5.286\pm0.001$&6.538&6.509$^{+0.16}_{-0.2}$&26.2 $\pm$ 0.2&-16.8&21.2$^{+1.6}_{-1.5}$\\ \hline
1423.16	&\parbox{1cm}{\centering C14215A1\\C14215A2}&215.928929	&24.072848	&9847.8 $\pm$ 0.6$^*$	&8.5 &189$\pm$25&5.5$\pm$0.7&3.0$^s$&Q1&$7.101^{+0.000}_{-0.001}$&6.664&[1.196,7.379]&27.5 $\pm$ 0.3&-19.1&1.38$^{+0.02}_{-0.02}$\\
\hline
1423.37	&\parbox{1cm}{\centering miki14D\\miki14D2}&215.93618	&24.074682	&6665.2 $\pm$ 0.8 &3.5 &3.71$\pm$2.39&0.3$\pm$0.1&0.4&Q1&$4.483^{+0.001}_{-0.002}$&4.537&4.421$^{+0.48}_{-0.58}$&26.0 $\pm$ 0.1&-19.4&2.06$^{+0.06}_{-0.05}$\\ \hline
370.43	&\parbox{1cm}{\centering A37017B1\\A37017B2}&39.966302	&$-$1.587095 &9253.6 $\pm$ 0.7	&4.7 &97.0$\pm$29.8&0.7$\pm$0.2&1.7&Q1&$6.612\pm0.001$&10.563&6.629$^{+0.16}_{-0.20}$&27.9 $\pm$ 0.3&-16.6&8.40$^{+0.02}_{-0.03}$\\ \hline
0416.89   &\parbox{1cm}{\centering 041615B1}&64.048668   &$-$24.082184   &7587.3 $\pm$ 0.7$^{s}$   &6.4   &35.3$\pm$16.0&0.3$\pm$0.1&4.0&Q1&$5.241\pm0.001$&3.027&3.055$^{+0.47}_{-0.37}$&27.7 $\pm$ 0.1&-17.9&2.20$^{+0.05}_{-0.05}$\\ \hline
2744.116	&\parbox{1cm}{\centering 274415B1}&3.592285	&$-$30.409911	&8632.9$^{8,*}$	&0.3 &224$\pm$60&1.2$\pm$0.3&$^*$&Q1&$6.101^{+0.001}_{-0.000}$&7.579&7.310$^{+0.51}_{-1.02}$&28.0 $\pm$ 0.2&-17.1&4.3$^{+0.5}_{-0.4}$\\ \hline
0717.59    &\parbox{1cm}{\centering 071715B1}&109.37792   &37.742851   &8068.3 $\pm$ 0.5$^s$  &3.4   &0.484$\pm$0.182&0.3$\pm$0.1&2.0&Q1&$5.637\pm0.001$&0.817&5.238$^{+0.15}_{-4.45}$&23.533 $\pm$ 0.008&-22.3&1.95$^{+0.01}_{-0.01}$\\ \hline
0744.17	&\parbox{1cm}{\centering M744D\_D1\\M744D\_D2}&116.202128	&39.44821	&8166.5 $\pm$ 0.7	&5.5 &25.9$\pm$13.1&0.2$\pm$0.1&4.0&Q1&$5.718^{+0.000}_{-0.001}$&0.835&5.216$^{+0.98}_{-4.36}$&27.4 $\pm$ 0.5&-18.9&1.207$^{+0.002}_{-0.003}$\\ \hline
1149.51    &\parbox{1cm}{\centering 114915A1}&177.413006 &22.4188617 &9353.0 $\pm$ 0.4 &5.7 &17.1$\pm$4.6&1.0$\pm$0.3&2.0&Q1&$6.694^{+0.000}_{-0.001}$&7.081&7.072$^{+0.17}_{-0.18}$&25.07 $\pm$ 0.08 &-21.1&1.82$^{+0.06}_{-0.04}$\\ \hline
1149.67    &\parbox{1cm}{\centering 114916A1}&177.412021 &22.415777 &9265.1 $\pm$ 0.6$^3$ &3.8 &6.38$\pm$2.27&0.8$\pm$0.3&0.5&Q1&$6.621\pm0.001$&6.031&5.990$^{+0.07}_{-0.09}$&24.91 $\pm$ 0.02&-21.3&1.77$^{+0.04}_{-0.03}$\\ \hline

\end{tabular}
ID 1423.16 corresponds to a detection with continuum flux density measured from the CLASH catalog.\\
$^s$ \textrm{These are secondary objects.}\\
$^*$ \textrm{Indicates these emission lines are next to or on skylines. The centroid location errors and asymmetry values should not be trusted. N.B. The object in A2744 has an emission line covering many skylines and we were unable to get a formal uncertainty and asymmetry value.}\\
$^3$\textrm{Photometrically selected by \citet{huang2016spitzer}.}\\
$^6$\textrm{Spectroscopically observed by \citet{treu2015grism}.}\\
$^7$\textrm{Spectroscopically observed by \citet{hoag2019constraining}.}\\
$^8$\textrm{Spectroscopically confirmed by, but inconsistent with \cite{mahler2017strong}.}
\end{table}
\end{turnpage}
\clearpage

\begin{figure*}[!htbp]
  \includegraphics[width=\linewidth]{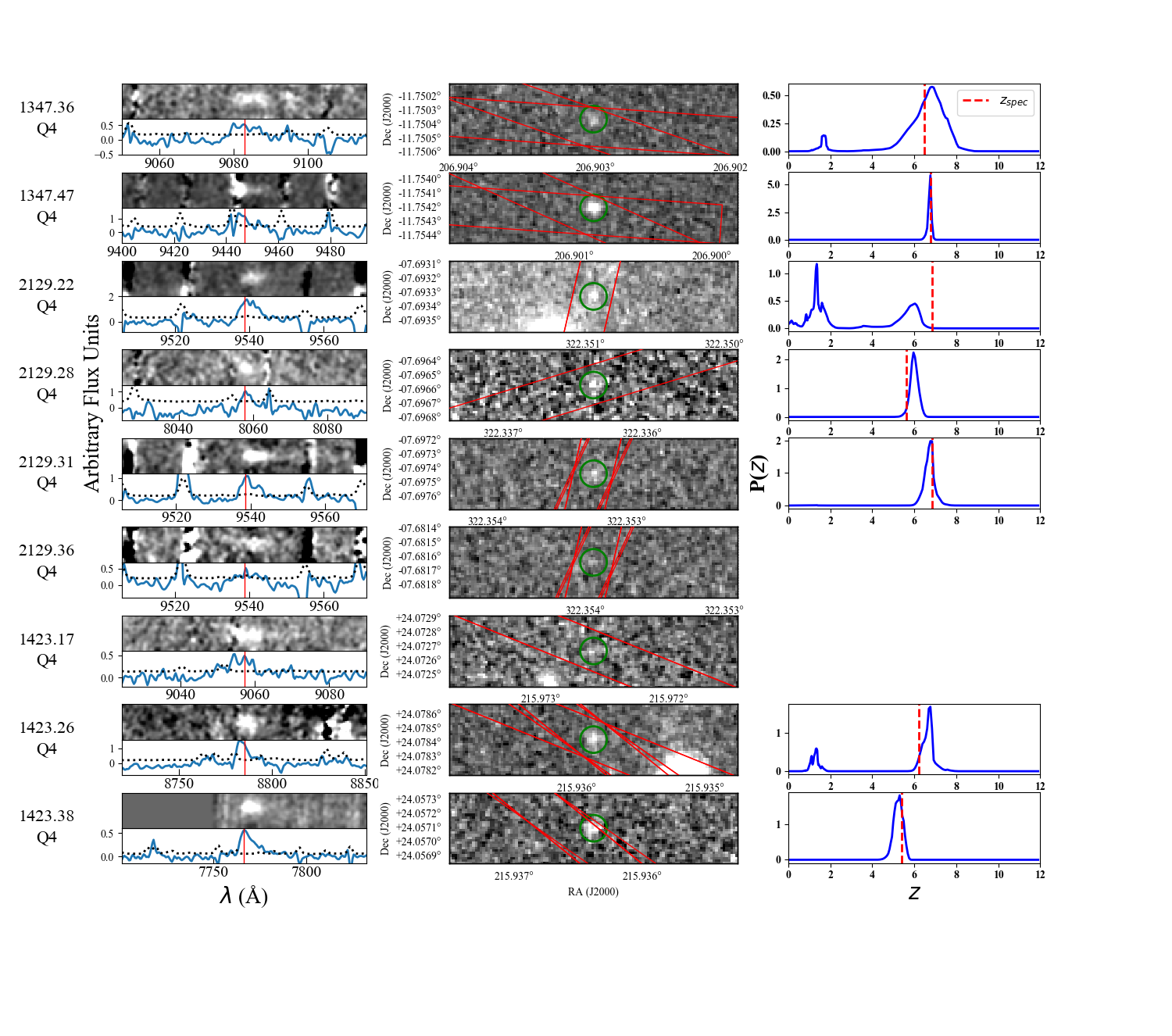}
  \caption{Left: Stacked 2D and 1D spectra of our detections. The 2D spectra are cutouts, centered on the emission line. The spectra have been smoothed to improve visibility of the emission line. The red line in the 1D spectra indicates the spectral centroid of the detection. The noise spectrum is also shown as a black dotted line. Center: A slits-on image in F160W of our candidate LAEs. If an object was observed on multiple masks, multiple slits are shown. Some slits are slightly offset as they are aligned with respect to the Subaru images. The green circle indicates the location of the candidate LAE from the photometric catalog. Right: The P($z$) distribution and spectroscopic redshift. The four CLASH detections (2129.36, 1423.17, 0416.17, and 1423.16) have crude P($z$) information described in Section 2.6 and are therefore not shown. The ID and quality of the detection is also displayed on the left. Objects marked with an asterisk are secondary objects. The figures on the following pages are ordered by quality of detection, from our most confident detections to our least confident detections, and then by cluster and ID.}
\end{figure*}

\begin{figure*}[!htbp]
  \ContinuedFloat
  \includegraphics[width=\textwidth]{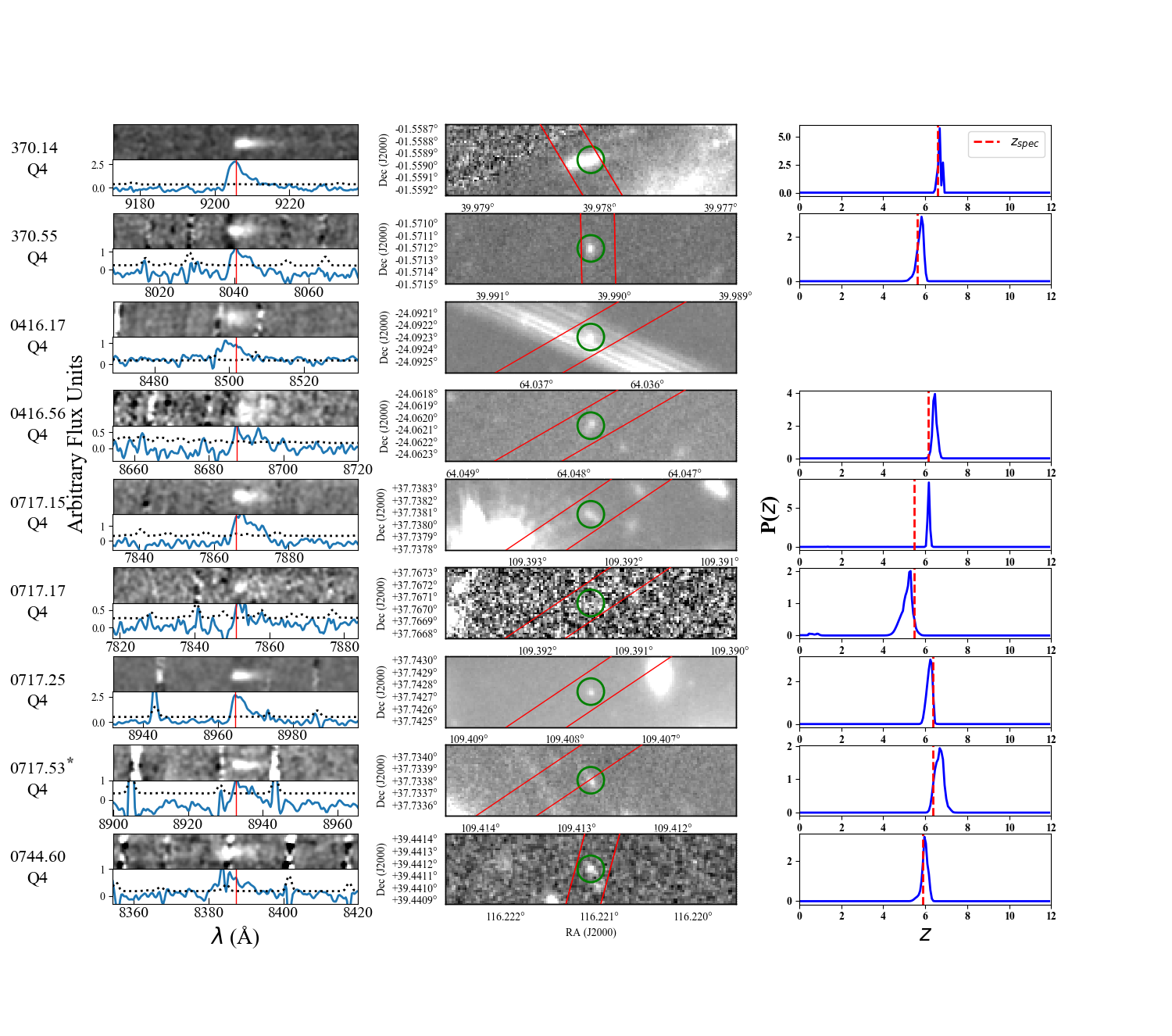}
  \caption{}
\end{figure*}

\begin{figure*}[!htbp]
  \ContinuedFloat
  \includegraphics[width=\textwidth]{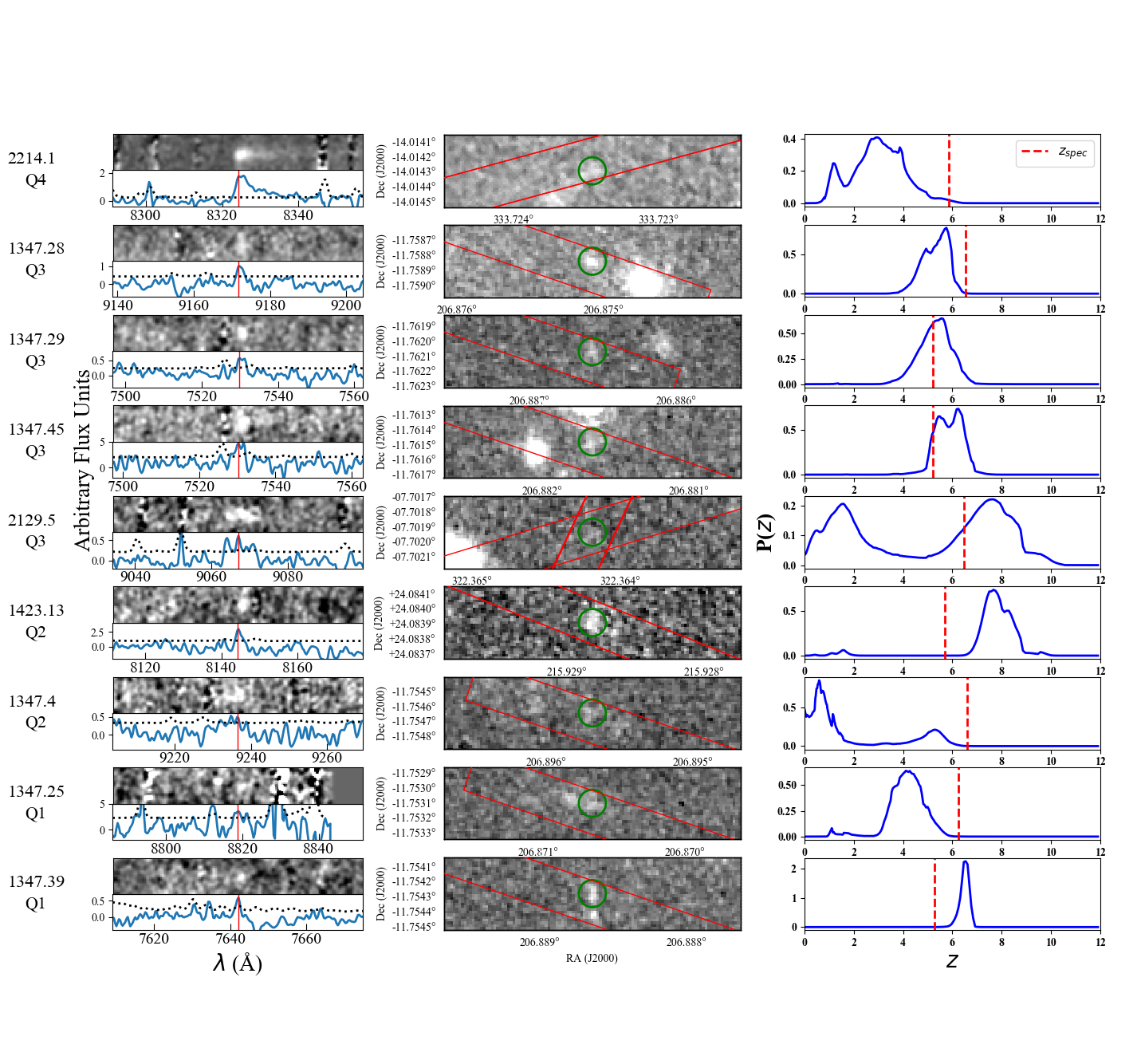}
  \caption{}
\end{figure*}

\begin{figure*}[!htbp]
  \ContinuedFloat
  \includegraphics[width=\textwidth]{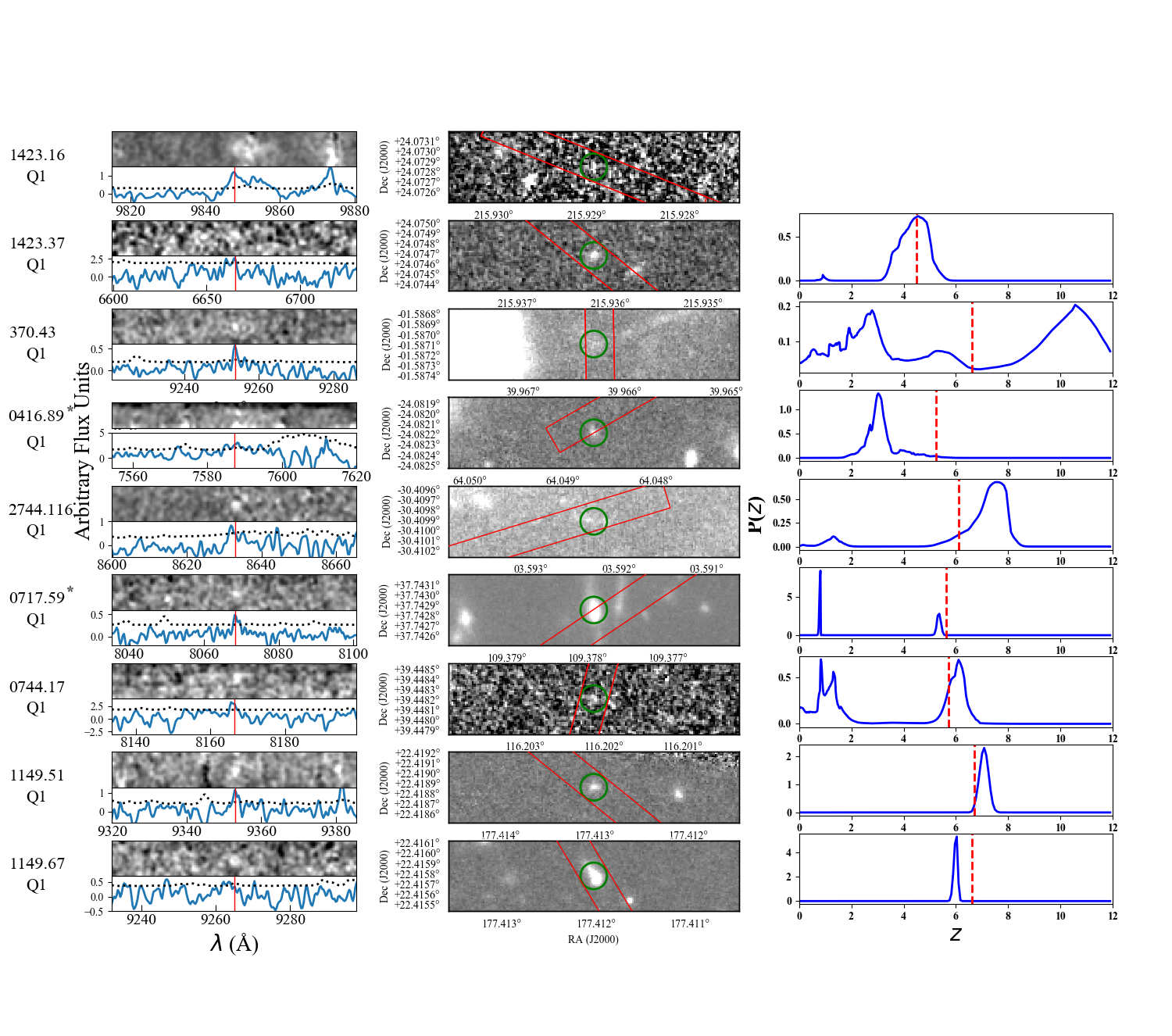}
  \caption{}
\end{figure*}
\clearpage

\subsection*{C. Multiply Imaged Ly$\alpha$ Emitters}
Five of our Ly$\alpha$ emitters are multiply imaged systems, one triply-imaged and one doubly-imaged. The three objects in MACS2129 (2129.22, 2129.31, and 2129.36) marked with $^1$ in Table 3 are three images of a LAE published in \citet{huang2016detection}. Two Ly$\alpha$ emitters in RXJ1347 (1347.29 and 1348.45) have Q3 detections at $\sim$7530Å. Both images have similar colors, and, more importantly, have the same $z_{spec}$, therefore, are likely multiple images.

\begin{table}[H]
\begin{center}
\caption{ASTRODEEP \& CLASH $m_{\textrm{F160W}}$ Offsets}
\begin{tabular}{| c c c|}
\hline \hline
$m_{\textrm{F160W}}$ & \parbox{2cm}{\centering Median Offset \\ (mag.)} & \parbox{4cm}{\centering Normalized Median \\ Absolute Deviation \\(mag.)} \\ \hline
20-21&0.07&0.09\\
21-22&0.08&0.09\\
22-23&0.08&0.11\\
23-24&0.05&0.13\\
24-25&0.05&0.15\\
25-26&0.04&0.19\\
26-27&0.04&0.27\\

\hline
\end{tabular}			
\end{center}
\textrm{F160W apparent magnitude ($m_{\textrm{F160W}}$) median offsets and normalized median average deviations for ASTRODEEP and CLASH objects.}
\end{table}

\end{document}